# Many-Objective Software Remodularization using NSGA-III

WIEM MKAOUER, University of Michigan, Dearborn
MAROUANE KESSENTINI, University of Michigan, Dearborn
ADNAN SHAOUT, University of Michigan, Dearborn
PATRICE KONTCHOU, University of Michigan, Dearborn
SLIM BECHIKH, University of Michigan, Dearborn
KALYANMOY DEB, Michigan State University
ALI OUNI, University of Michigan, Dearborn

**Abstract.** Software systems nowadays are complex and difficult to maintain due to continuous changes and bad design choices. To handle the complexity of systems, software products are, in general, decomposed in terms of packages/modules containing classes that are dependent. However, it is challenging to automatically remodularize systems to improve their maintainability. The majority of existing remodularization work mainly satisfy one objective which is improving the structure of packages by optimizing coupling and cohesion. In addition, most of existing studies are limited to only few operation types such as move class and split packages. Many other objectives, such as the design semantics, reducing the number of changes and maximizing the consistency with development change history, are important to improve the quality of the software by remodularizing it. In this paper, we propose a novel many-objective search-based approach using NSGA-III. The process aims at finding the optimal remodularization solutions that improve the structure of packages, minimize the number of changes, preserve semantics coherence, and re-use the history of changes. We evaluate the efficiency of our approach using four different open-source systems and one automotive industry project, provided by our industrial partner, through a quantitative and qualitative study conducted with software engineers.

Categories and Subject Descriptors: D.2 **[Software Engineering]**.

## 1. INTRODUCTION

Large software systems evolve and become complex quickly, fault-prone and difficult to maintain [Mitchell 2006]. In fact, most of the changes during the evolution of systems such as introducing new features or fixing bugs are conducted, in general, within strict deadlines. As a consequence, these code changes can have a negative impact on the quality of systems design such as the distribution of the classes in packages. To address this issue, one of the widely used techniques is software remodularization, called also software restructuring, which improves the existing decomposition of systems.

There has been much work on different techniques and tools for software remodularization [Wiggerts 1997; Mancoridis et al. 1998; Anquetil and Lethbridge 1999; Mitchell and Mancoridis 2006; Maqbool and Babri 2007; Mitchell and Mancoridis 2008; Abdeen et al. 2009; Shtern and Azerpos 2009; Bavota et al. 2010]. Most of these studies addressed the problem of clustering by finding the best decomposition of a system in terms of modules and not by improving existing modularizations. In both categories, cohesion and coupling are the main metrics used to improve the quality of existing packages (e.g. modules) by determining which classes need to be grouped in a package. In this paper, we focus on restructuring software design and not on the decomposition of systems to generate an initial coherent object oriented design.

The majority of existing contributions have formulated the restructuring problem as a single-objective problem where the goal is to improve the cohesion and coupling of packages [Abdeen et al. 2009; Bavota et al. 2010; Praditwong et al. 2011; Bavota et al. 2013]. Even though most of the existing approaches are powerful enough to provide remodularization solutions, some issues still need to be addressed.

One of the most important issues is the semantic coherence of the design. The restructured program could improve structural metrics but become semantically incoherent. In this case, the design will become difficult to understand since classes are placed in wrong packages to improve the structure in terms of cohesion and coupling. Also, the number of code changes is not considered when suggesting remodularization solutions; the only aim is to improve the structure of packages independently of the cost of code changes.

However, in a real-world scenarios, developers prefer, in general, remodularization solutions that improve the structure with a minimum number of changes. It is important to minimize code changes to help developers in understanding the design after applying suggested changes.

Existing remodularization studies are also limited to few types of changes mainly *move class* and *split packages* [Abdeen et al. 2009; Bavota et al. 2010; Bavota et al. 2012; Bavota et al. 2013; Abdeen et al. 2013]. However, refactoring at the class and method levels can improve remodularization solutions such as by moving methods between classes located in different packages. The use of development history can be an efficient aid when proposing remodularization solutions [Palomba et al. 2013]. For example, packages that were extensively modified in the past may have a high probability of being also changed in the future. Moreover, the packages to modify can be similar to some patterns that can be found in the development history, thus, developers can easily recognize and adapt them.

In this paper, we propose a many-objective search-based approach [Harman et al. 2012; Deb and Jain 2014] to address the above-mentioned limitations. Search-based software engineering is suitable for the software remodularization problem since the goal is to find the best sequence of operations that can lead to a better remodularized system. The number of combinations to explore is high, leading to a huge and complex search space. Our many-objective search-based software engineering approach aims at finding the remodularization solution that: 1) Improve the structure of packages by optimizing some metrics such as number of classes per package, number of packages, coupling and cohesion; 2) Improve the semantic coherence of the restructured program. We combine two heuristics to estimate the semantic proximity between packages when moving elements between them (vocabulary similarity, and dependencies between extracted classes from call graphs) and some semantic/syntactic heuristics depending on the change type; 3) Minimize code changes. Compared to existing remodularization studies, we consider new changes that can be related to the package, class and method levels; and 4) Maximize the consistency with development change history. To better guide the search process, recorded code changes that are applied in the past in similar contexts are considered. We evaluate if similar changes are applied in previous versions of the packages that will be modified by the suggested remodularization solution.

The number of objectives to consider in our problem formulation is high (more than three objectives); such problems are termed many-objective. In this context, the use of traditional multi-objective techniques, e.g., NSGA-II [Deb et al. 2002], widely used in Search-Based Software Engineering (SBSE) [Harman and Tratt 2007; Ferrucci et al. 2013], is clearly not sufficient like in our case for the problem of software remodularization. There is a growing need for SBSE approaches that address software engineering problems where a large number of objectives are to be optimized. Recent work in optimization has proposed several solution approaches to tackle many-objective optimization problems [Jaimes et al. 2009; Deb and Jain 2012; Jain and Deb 2013; Deb and Jain 2014] using e.g., objective reduction, new preference ordering relations and decomposition. However, these techniques have not yet been widely explored in SBSE [Harman et al. 2012]. To the best of our knowledge and based on recent SBSE surveys [Harman 2013], only one work exists proposed by [Sayyad et al. 2013a; Sayyad et al. 2013b] that uses a many-objective approach, IBEA (Indicator-Based Evolutionary Algorithm) [di Pierro et al. 2007], to address the problem of software product line creation. However, the number of considered objectives is limited to five.

We propose a scalable search-based software engineering approach based on NSGA-III [Deb and Jain 2014] where there are seven different objectives to optimize. Thus, in our approach, automated remodularization solutions will be evaluated using a set of seven measures as described above. The basic framework remains similar to the original NSGA-II algorithm, with significant changes in its selection mechanism. This paper represents one of the first real-world applications of NSGA-III and the first scalable work that supports the use of seven objectives to address and improve software remodularization.

We evaluated our approach on four open source systems and one industrial system provided by our industrial partner Ford Motor Company. We report the results on the efficiency and effectiveness of our approach, compared to the state of the art remodularization approaches [Abdeen et al. 2009; Abdeen et al. 2013; Bavota et al. 2013]. Our results indicate that our approach significantly outperforms, in average, existing approaches in terms of improving the structure, reducing the number code changes, and semantics preservation.

The primary contributions of this paper can be summarized as follows:
1. The paper introduces a novel formulation of the remodularization problem as a many-objective problem that considers several objectives such as structural improvement, semantic coherence, number of changes and consistency with history of changes.
2. We consider in the paper the use of new operations, comparing to existing remodularization studies including move method, extract class and merge packages.
3. The paper reports the results of an empirical study of our many-objective technique compared to different existing approaches. The obtained results provide evidence to support the claim that our proposal is, in average, more efficient than existing techniques based on a benchmark of four large open source systems and one industrial project.
4. The qualitative evaluation of the results by software engineers at Ford Motor Company and also graduate students confirms the usefulness the suggested remodularization solutions.

The remainder of this paper is structured as follows: Section 2 is dedicated to the background needed to understand our approach and to the remodularization challenges. In Section 3, we describe how software remodularization is formulated as many-objective optimization problem and explain how we adapted the NSGA-III algorithm to our problem in Section 4. Section 5 presents and discusses the evaluation results on several medium and large size projects. We discuss in section 6 the threats to validity related to our experiments. Related work is outlined in section 7. Section 8 concludes the paper and suggests future research directions.

## 2. SOFTWARE REMODULARIZATION: BACKROUND AND MOTIVATING EXAMPLE

### 2.1 Background

Large systems such as automotive industry applications have to run and evolve over decades. As described by the law of [Lehman 1980], most of industrial systems must evolve and the design is, in general, extended far away the initial structure. Thus, it is mandatory to restructure the program design to reduce the cost of possible future evolutions. To this end, software remodularization is an important component in software maintenance activities.

Object oriented software modularization consists of regrouping a set of classes $C$ in terms of packages $P$. Thus, each package $P$ contains a set of classes. Several types of dependencies between packages can be found in the literature [Wiggerts 1997; Anquetil and Lethbridge 1999; Seng et al. 2005; Praditwong et al. 2011; Bavota et al. 2012; Abdeen et al. 2013]. In this paper, we use the definition of dependencies between packages defined in [Praditwong et al. 2011]. Two main types of dependencies are described: 1) intra-edges dependencies and 2- inter-edges dependencies. The intra-edges include all types of internal dependencies between classes in the same package such as method call, class reference and inheritance. 2) The inter-edges include external dependencies between classes that are not in the same package. As illustrated in Figure 1, the system includes 2 packages, 3 intra-edges such as ($c_1$, $c_3$) and 2 inter-edges such as ($c_3$, $c_4$) for package $P_1$.

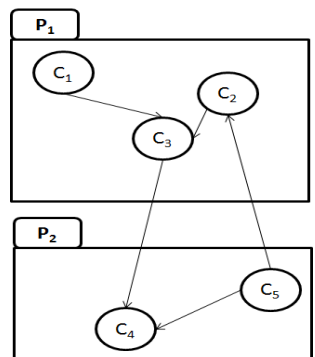

**Fig. 1.** The dependency graph including two packages, 3 intra-edges and 2 inter-edges.

Most of existing approaches are based on the use of cohesion and coupling to evaluate remodularization solutions.

The best solutions are those that maximize cohesion and minimize coupling. Cohesion of packages is, in general, defined by the number of intra-edges of a package and coupling as the number of inter-edges of a package. Packages by their very natures should be highly cohesive in many cases. In programming languages developers need to import packages in order to have visibility to the classes inside of them. This provides a natural check and balance that the package structure makes syntactic and semantic sense. If the packages were hard to import, users of those packages would either change them, or request that structural changes are made. Our goal is to try to improve this structuring to make the system easier to maintain. Even though most of the existing approaches are powerful enough to provide remodularization solutions, some open issues need to be targeted to provide an efficient and fully automated remodularization. In the next section, we illustrate these different limitations mentioned in the introduction section.

## 2.2 Motivating Example

To illustrate some of the issues mentioned in the first section, Figure 2 shows a concrete example extracted from GanttProject v1.10.2, a well-known Java open-source project management software. We consider a design fragment containing four packages *net.sourceforge.ganttproject*, *net.sourceforge.ganttproject.document*, *net.sourceforge.ganttproject.gui,* and *net.sourceforge.ganttproject.task*. The largest package in GanttProject v1.10.2 is *net.sourceforge.ganttproject* including more than 40 classes, comparing to all other packages, implementing several features in one package. All the twelve software engineers that we asked in our experiments agreed that *net.sourceforge.ganttproject* is a large package that monopolizes the behavior of a large part of the system.

We consider an example of a remodularization solution that consists of moving some classes from the package *net.sourceforge.ganttproject* to the package *net.sourceforge.ganttproject.document*. This operation can improve the modularization quality by reducing the number of classes/functionalities of the package *net.sourceforge.ganttproject*. However, from the semantics coherence standpoint, all the classes of *net.sourceforge.ganttproject.document* are different from those implemented in *net.sourceforge.ganttproject* since they implement a feature to open streams to a project file. Based on semantic and structural information, using respectively a vocabulary-based similarity, and cohesion/coupling, many other target packages are possible including *net.sourceforge.ganttproject.gui,* and *net.sourceforge.ganttproject.task*. These two packages have almost the same structure based on metrics such as number of classes and their semantic similarity is close to *net.sourceforge.ganttproject* using a vocabulary-based measure, or cohesion and coupling. On the other hand, from previous versions of GanttProject, we recorded that there are some classes (e.g. *TaskManagerImpl*) that have been moved from the package *net.sourceforge.ganttproject* to the package *net.sourceforge.ganttproject.task*. As a consequence, moving from the package *net.sourceforge.ganttproject* to the package *net.sourceforge.ganttproject.task* has higher correctness probability than moving classes between the remaining packages. Thus, some classes can be moved between these two packages such as *GanttTask, GanttTaskPropertiesBean* and *GanttTaskRelationship*. The direction of class movement should be taken into account while analyzing the history of changes to find similarities.

Based on these observations, we believe that it is important to consider additional objectives rather than using only structural metrics to improve the automation of software remodularization. However, in most of the existing remodularization work, semantic coherence, code changes, and development history are not considered. Thus, the remodularization process needs a manual inspection by the user to evaluate the meaningfulness/feasibility of proposed changes that mainly improve structural metrics. The inspection aims at verifying if these changes could produce semantic incoherence in the program design. For large systems, this manual inspection is complex, time-consuming and error-prone. Improving the packages structure, minimizing semantic incoherencies, reducing code changes, and keeping consistent with development change history may be conflicting. In some cases, improving the program modularization could provide a design that does not make sense semantically or could change radically the initial design. For these reasons, a good remodularization strategy needs to find a compromise between all of these objectives. In addition, moving classes and splitting packages are not enough code changes to improve the remodularization of systems. These observations are at the origin of the work described in this paper.

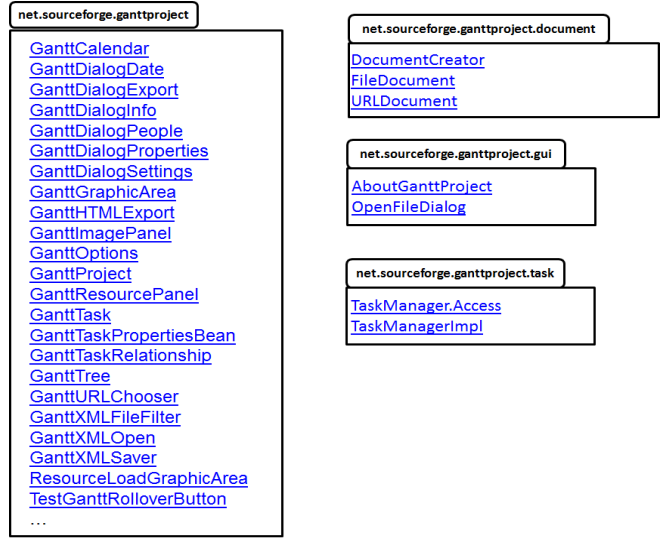

**Fig. 2.** Motivating example extracted from GanttProject v1.10.2.

## 3. SOFTWARE REMODULARIZATION: A MANY-OBJECTIVE PROBLEM

We describe in this section how many-objective techniques can be adapted to the software remodularization problem. Initially, we start by describing many-objective techniques, then we illustrate how the software remodularization problem can be considered as a many-objective one.

### 3.1 Many-Objective Search-based Software Engineering

Recently many-objective optimization has attracted much attention in Evolutionary Multi-objective Optimization (EMO) which is one of the most active research areas in evolutionary computation [Deb and Jain 2012]. By definition, a many-objective problem is multi-objective one but with a high number of objectives $M$, i.e., $M > 3$. Analytically, it could be stated as follows:

$$\begin{cases} Min f(x) = [f_1(x), f_2(x), ..., f_M(x)]^T, & M > 3 \\ g_j(x) \geq 0 & j = 1,...,P; \\ h_k(x) = 0 & k = 1,...,Q; \\ x_i^L \leq x_i \leq x_i^U & i = 1,...,n \end{cases} \quad (1)$$

where $M$ is the number of objective functions and is *strictly greater* than 3, $P$ is the number of inequality constraints, $Q$ is the number of equality constraints, $x_i^L$ and $x_i^U$ correspond to the lower and upper bounds of the decision variable $x_i$ (i.e., $i^{th}$ component of $x$). A solution $x$ satisfying the $(P+Q)$ constraints is said to be feasible and the set of all feasible solutions defines the feasible search space denoted by $\Omega$.

In this formulation, all the considered objectives are to be minimized, since maximization can be easily turned to minimization based on the duality principle. Over the two past decades, several Multi-Objective Evolutionary Algorithms (MOEAs) have been proposed with the hope to work with any number of objectives $M$. Unfortunately, it has been demonstrated that most MOEAs are ineffective in handling such type of problems. For example, NSGA-II, which is one of the most popular MOEAs, compares solutions based on their non-domination ranks. Solutions with best ranks are emphasized in order to converge to the Pareto front. When $M > 3$, only the first rank may be assigned to every solution as almost all population individuals become non-dominated with each other [Khare 2003; Wang and Yao 2010]. Without a variety of ranks, NSGA-II cannot keep the adequate search pressure in high dimensional objective spaces.

Software engineering problems require a search for a solution which balances multiple objectives and constraints to achieve near optimal or optimal results. This search can be fastidious and requires a labor-intensive human activity. SBSE has provided new ways, based on heuristics, transforming software engineering problems from human-based search to machine-based search techniques. Thus, the use of heuristics can guide the automated search and avoid the tedious human-in-the-loop search activities. However, currently existing techniques lack scalability to meet the demands of high dimensional software engineering solutions. According to a recent survey by Harman [Harman 2013], most software engineering problems are naturally multi-objective. However, they are being mostly handled from a mono-objective perspective. Multi-objective optimization techniques have been proposed in a few works [Harman and Tratt 2007; Durillo et al. 2009; Ouni et al. 2013a; Ouni et al. 2013b] for such problems and they satisfy up to five objectives. However, as with any other practical domain, most software engineering problems involve optimizing more than this number of objectives. Thus, more scalable SBSE approaches will be beneficial to handle rich objective spaces.

The difficulty faced when solving a many-objective problem could be summarized as follows. Firstly, many solutions become equivalent according to the Pareto dominance theory which induces the dramatically deterioration of the search process and so the inability to converge towards the Pareto front; consequently, the MOEA behaviour becomes very similar to the one of a random search. Secondly, high dimensional problems require, for their resolution, a large number of solutions to maximize the coverage of the search space. As an illustrative example, in order to reach a suitable approximation of the Pareto front of an optimization problem with 4, 5 and 7 objectives, the number of required non-dominated solutions is about 62 500, 1 953 125 and 1 708 984 375 respectively [Jaimes et al. 2009]; which makes the decision making task very difficult. Thirdly, the higher is the objective space cardinality, the harder is to locate promising search directions. Furthermore, determining the crowding distance of each solution in a large population for many fitness functions is eventually a complex calculation; thus, measuring the population diversity is definitely expensive from a computational perspective [Bechikh et al. 2010; Jaimes et al. 2014]. Moreover, in a large search space, the limited number of solutions increases their distances from each other. This will decrease the similarity between parents and offsprings; thereby making the inefficiency of recombination operation in producing promising solutions. Finally, from user-visualization view, plotting the Pareto-equivalent solutions becomes a complicated task, therefore making the interpretation of the MOEA's results more difficult.

Recently, researchers have proposed several approaches to tackle many-objective optimization problems. Firstly, we state the o*bjective reduction approach*. It mainly look for the minimal subset of conflicting objectives. The objective reduction approach attempts initially examine the degree of conflict among objectives in order to eliminate objectives that do not construct the Pareto- front [Saxena 2013]. Regardless of the number of objectives, finding objectives reduction opportunities in a problem has a favorable impact on the search efficiency, computational cost, and decision making. Although this technique has solved benchmark problems involving up to 20 objectives, its applicability in real world setting is not straightforward and it remains to be investigated since most objectives are usually in conflict with each other [Ó Cinnéide et al. 2012]. Secondly, we note the *incorporation of decision maker's preferences*. With the increase of objectives, the Pareto optimal approximation involves the investigation of a large number of Pareto-equivalent solutions. Consequently, the numerous variety of solutions makes the choice of the preferred alternative very hard for the human decision maker (DM). More practically, DMs are not usually interested in the whole Pareto front rather than a portion of it that best fits their preferences, called the Region of Interest (ROI). The main idea is to incorporate the DM's preferences in the search space in order to distinguish between Pareto equivalent solutions that have the ability to evolve towards the ROI on problems involving more than 3 objectives [Ben Said et al. 2010; Bechikh et al. 2011]. Preference-based MOEAs have given many interesting results when addressing concrete problems in several engineering fields including software design by incorporating the designer preferences [Jaimes et al. 2011; Siegmund et al. 2012; Mkaouer et al. 2013]. Thirdly, the *new preference ordering relations* is an alternative approach that takes into account additional information such as the rank of the particular solution regarding the different objectives and the related population [di Pierro 2007] in order to overcome the inability of differentiating between solutions with the increased of the number of objectives; however these methods do not necessarily agree with to the DMs preferences. Fourthly, the *decomposition* technique consists in decomposing the problem into several sub-problems that can be solved simultaneously by using evolutionary algorithms' parallel search capability. The most reputable decomposition-based evolutionary algorithm is MOEA/D [Zhang and Li 2007]. Finally, the u*se of a* *predefined multiple targeted search*.is inspired by preference-based MOEAs and the decomposition approach. Recently, [Jain and Deb 2013] and [Wang et al. 2013] have proposed a new idea that involves guiding the population during the optimization process based on multiple predefined targets (e.g., reference points, reference direction) in the objective space. This idea has demonstrated very promising results on MOPs involving up to 15 objectives. Table 1 illustrates a summary of existing many-objective techniques.

**Table 1.** Summary of many-objective approaches.

| Approach | Basic idea | Example algorithms | #objectives | Applications |
|---|---|---|---|---|
| Objective reduction | Find the minimal subset of conflicting objectives, then eliminate the objectives that are not essential to describe the Pareto optimal front. | 1)PCA-NSGA-II [Deb and Saxena 2006]<br>2) PCSEA [Singh et al. 2011] | 10<br><br>20 | 1) Not yet<br>2) Water resource problem |
| Incorporating decision maker's preferences | Exploit DM's preferences in order to differentiate between Pareto equivalent solutions so that we can direct the search towards the region of interest instead of the whole front. | 1) r-NSGA-II [Ben Said et al. 2010]<br>2) PBEA [Thiele et al. 2009]<br>3) R-NSGA-II [Deb et al. 2006] | 10<br><br>10<br><br>10 | 1) Payment scheduling problem<br>2) Not yet<br>3) Welded beam design problem |
| New preference ordering relations | Propose alternative preference relations that are different from the Pareto dominance. | 1) Preference Order Ranking–based algorithm [di Pierro 2007]<br>2) Ranking dominance-based algorithm [Kukkonen and Lampinen 2007]<br>3) IBEA [Zitzler and Künzli 2004]<br>4) HypE [Bader and Zitzler 2011] | 8<br><br><br>10<br><br><br><br>5<br><br>20 | 1) Water distribution problem<br><br>2) Not yet<br>3) Software product line management<br>4) Not yet |
| Decomposition | Decompose the problem into several sub-problems and then solve these sub-problems simultaneously by exploiting the parallel search ability of EAs. | 1) MOEA/D [Zhang and Li 2007] | 5 | 1) Not yet |
| Use of a predefined multiple targeted search | Guide the population during the optimization process based on multiple predefined targets (e.g., reference points) in the objective space. | 1) PICEA [Wang et al. 2013]<br>2) NSGA-III [Deb and Jain 2014] | 10<br><br>15 | 1) Crash-worthiness Design of Vehicles 2) Not yet |

We investigate, in this paper, the applicability of many-objective techniques for the software remodularization problem where up-to 7 objectives are considered to evaluate remodularization suggestions.

### 3.2 Many-Objective Software Remodularization: Overview

*3.2.1   Approach Overview*

Our approach aims at exploring a huge search space to find a set of remodularization solutions on the Pareto front such that in order to optimize any objective further will result in sub-optimizing one or more additional objectives. These remodularization solutions are a sequence of change operations, to restructure packages. The search space is determined not only by the number of possible change combinations, but also by the order in which they are applied. A heuristic-based optimization method is used to generate remodularization solutions. We have 7 objectives to optimize: 1) minimize the number of classes per package; 2) minimize the number of packages; 3) maximize package cohesion; 4) minimize package coupling; 5) minimize the number of semantic errors by preserving the way classes are semantically grouped and connected together; 6) minimize code changes needed to apply remodularization solution; and 7) maximize the consistency with development change history. We consider the remodularization task as a many-objective optimization problem instead of a single-objective one using the new many-objective non-dominated sorting genetic algorithm (NSGA-III) that will be described in Section 4.

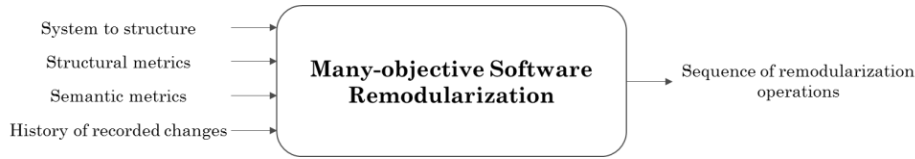

**Fig. 3.** Approach overview.

The general structure of our approach is sketched in Fig. 3. It takes as input the source code of the program to be restructured, a list of possible remodularization operations (ROs) that can be applied, a set of semantic and structural measures, and a history of applied changes to previous versions of the system. Our approach generates as output the optimal sequence of operations, selected from a list of possible ones that improve the structure of packages, minimize code changes needed to apply the remodularization solution, preserve the semantics coherence, and maximize the consistency with development change history. In the following, we describe the formal formulation of these different remodularization objectives to optimize.

*3.2.2  Remodularization Objectives*

We describe in this section the 7 objectives to optimize in our many-objective adaptation to the software remodularization problem.

**Structure.** Four conflicting objectives are related to improving the structure of packages: 1) *number of classes per package* (to minimize); 2) *number of packages in the system* (to minimize); 3) *cohesion* (to maximize); it corresponds to the number of intra-edges (calls between classes in the same package) as described in Section 2; and 4) *coupling* (to minimize); it corresponds to the number of inter-edges (class between classes in different packages).

**Number of code changes**. Table 2 describes the types of ROs that are considered by our approach: *Move method, Extract class, Move class, Merge packages,* and *Extract/Split package*. Existing remodularization studies are limited to only two operation types: *move class* and *split/extract package*. We believe that these two operations are not enough to generate good remodularization solutions. In fact, sometimes only part of the class should be moved to another package (e.g. methods) and not the whole class. To apply a remodularization operation we need to specify which *actors*, *i.e.*, code fragments, are involved in this operation and which *roles* they play when performing the change. As illustrated in Table 2, an actor can be a *package*, *class*, or *method* and we specify for each operation the involved actors and their roles. It is important to minimize the number of suggested operations in the remodularization solution since the designer can have some preferences regarding the percentage of deviation with the initial program modularization. In addition, most of developers prefer solutions that minimize the number of changes applied to their design [Ouni et al. 2012a].

**Table 2.** Types of remodularization operation.

| Type of the operation | Actors | Roles |
|---|---|---|
| Move method | class | source class, target class |
|  | method | moved method |
| Extract class | class | source class, new class |
|  | method | moved methods |
| Move class | package | source package, target package |
|  | class | moved class |
| Merge packages | package | source package, target package |
| Extract/Split package | package | source package, target package |
|  | class | moved class |

**Similarity with history of code changes.** The idea is to encourage the use of ROs that are similar to those applied to the same code fragments (packages) in the past. To calculate the similarity score between a proposed operation and a recorded code change, we use the following function:

$$SimilarityHistory(RO) = \sum_{j=1}^{n} w_j \qquad (2)$$

where *n* is the number of recorded operations applied to the system in the past, and $w_j$ is a change weight that reflects the similarity between the suggested remodularization operation (RO) and the recorded code change *j*. The weight $w_j$ is computed as described in Table 3.

Table 3. Similarity scores between remodularization operations applied to similar code fragments.

|  | Move method | Extract class | Move class | Merge packages | Extract/Split package |
|---|---|---|---|---|---|
| Move method | $w_j = 2$ | $w_j = 1$ | $w_j = 1$ | $w_j = 0$ | $w_j = 0$ |
| Extract class | $w_j = 1$ | $w_j = 2$ | $w_j = 0$ | $w_j = 0$ | $w_j = 0$ |
| Move class | $w_j = 1$ | $wj = 0$ | $w_j = 2$ | $w_j = 1$ | $w_j = 1$ |
| Merge packages | $w_j = 0$ | $wj = 0$ | $w_j = 1$ | $w_j = 2$ | $w_j = 0$ |
| Extract/Split package | $w_j = 0$ | $wj = 0$ | $w_j = 1$ | $w_j = 0$ | $w_j = 2$ |

**Semantics.** Most of ROs are simple to implement and it is almost trivial to show that they preserve the behaviour [Harman et al. 2002]. However, until now there is no consensual way to investigate whether a code change can preserve semantic coherence of the original program/design. To preserve the semantics design, some constraints should be satisfied to ensure the correctness of the applied operations. We distinguish between two kinds of constraints: structural constraints and semantic constraints. Structural constraints were extensively investigated in the literature. Opdyke, for example, defined in [Opdyke 1992] a set of pre and post-conditions for a large list of operations to ensure the structural consistence. Developers should check manually all code elements (packages, classes, methods and fields) related to the operation to inspect the semantic relationship between them. We formulate semantics constraints using different measures in which we describe the concepts from a perspective that helps in automating the remodularization task:

*Vocabulary-based similarity (VS)*

This kind of similarity is interesting to consider when moving methods, or classes or merging packages or extracting packages. For example, when a class has to be moved from one package to another, the operation would make sense if both code elements (source class and target packages) use similar vocabularies [Hamdi 2011; Ouni et al 2012a; Kim et al. 2013]. The vocabulary could be used as an indicator of the semantic similarity between different code elements that are involved when performing a remodularization operation. We start from the assumption that the vocabulary of an actor is borrowed from the domain terminology and therefore could be used to determine which part of the domain semantics is encoded by an actor. Thus, two code elements could be semantically similar if they use similar vocabularies.

The vocabulary could be extracted from the names of packages, classes, methods, fields, variables, parameters and types. Tokenisation is performed using the Camel Case Splitter [Hamdi 2011] which is one of the most used techniques in Software Maintenance tools for the preprocessing of identifiers. A more pertinent vocabulary can also be extracted from comments, commits information, and documentation. We calculate the semantic similarity between code elements using information retrieval-based techniques (e.g., cosine similarity). The following equation calculates the cosine similarity between two code elements. Each actor is represented as an n dimensional vector, where each dimension corresponds to a vocabulary term. The cosine of the angle between two vectors is considered as an indicator of similarity. Using cosine similarity, the conceptual similarity between two code elements *c1* and *c2* is determined as follows:

$$Sim(c1,c2) = \cos(\vec{c1}, \vec{c2}) = \frac{\vec{c1}.\vec{c2}}{\|\vec{c1}\| * \|\vec{c2}\|} = \frac{\sum_{i=1}^{n}(w_{i,1} * w_{i,2})}{\sqrt{\sum_{i=1}^{n}(w_{i,1})^2}\sqrt{\sum_{i=1}^{n}(w_{i,2})^2}} \in [0,1] \qquad (3)$$

where $\vec{c_1} = (w_{1,1},...,w_{n,1})$ is the term vector corresponding to actor *c1* and $\vec{c_2} = (w_{1,2},...,w_{n,2})$ is the term vector corresponding to *c2*. The weights $w_{i,j}$ can be computed using information retrieval based techniques such as the Term Frequency – Inverse Term Frequency (TF-IDF) method. We used a method similar to that described in [Ouni et al 2012b] to determine the vocabulary and represent the code elements as term vectors.

*Dependency-based similarity (DS)*

We approximate domain semantics closeness between code elements starting from their mutual dependencies. The intuition is that code elements that are strongly connected (i.e., having dependency links) are semantically related. As a consequence, ROs requiring semantic closeness between involved code elements are likely to be successful when these code elements are strongly connected. We consider two types of dependency links:

1) *Shared method calls (SMC)* that can be captured from call graphs derived from the whole program using CHA (Class Hierarchy Analysis) [Kim et al. 2013]. A call graph is a directed graph which represents the different calls (call in and call out) among all methods of the entire program. Nodes represent methods, and edges represent calls between these methods. CHA is a basic call graph that considers class hierarchy information, e.g, for a call *c.m(...)* assume that any *m(...)* is reachable that is declared in a super-type of the declared type of *c*. For a pair of code elements, shared calls are captured through this graph by identifying shared neighbours of nodes related to each actor. We consider both, shared call-out and shared call-in. The following equations are used to measure respectively the shared call-out and the shared call-in between two code elements $c_1$ and $c_2$ (two classes, for example). A shared method call is defined as the average of shared call-in and call-out.

$$sharedCallOut(c1,c2) = \frac{|callOut(c1) \cap callOut(c2)|}{|callOut(c1) \cup callOut(c2)|} \in [0,1] \qquad (4)$$

$$sharedCallIn(c1,c2) = \frac{|callIn(c1) \cap callIn(c2)|}{|callIn(c1) \cup callIn(c2)|} \in [0,1] \qquad (5)$$

2) *Shared field access (SFA)* can be calculated by capturing all field references that occur using static analysis to identify dependencies based on field accesses (read or modify). We assume that two software elements are semantically related if they read or modify the same fields. The rate of shared fields (read or modified) between two code elements $c_1$ and $c_2$ is calculated according to the following equation. In this equation, *fieldRW($c_i$)* computes the number of fields that may be read or modified by each method of the actor $c_i$. Thus, by applying a suitable static program analysis to the whole method body, all field references that occur could be easily computed.

$$sharedFieldsRW(c1,c2) = \frac{|fieldRW(c1) \cap fieldRW(c2)|}{|fieldRW(c1) \cup fieldRW(c2)|} \in [0,1] \qquad (6)$$

*Cohesion-based dependency (CD)*

The cohesion-based similarity that we propose for software remodularization is mainly used by the extract class, merge packages and extract package operations. It is defined to find a cohesive set of classes and methods to be moved to the newly extracted class or package. A new class or package can be extracted from a source class or package by moving a set of strongly related (cohesive) classes and methods from the original class or package to the new class or package. Extracting this set will improve the cohesion of the original package or class and minimize the coupling with the new package/class. Applying the *Extract Package/Class* or *Merge Packages* operation on a specific package/class will result in this class being split (or merged) into classes/packages. We need to calculate the semantic similarity between the elements in the original package/class to decide how to split or merge the original packages/classes.

We use vocabulary-based similarity and dependency-based similarity to find the cohesive set of code elements. For *move method*, the cohesion matrix is composed by the fields of the source method as lines, and the fields and methods of the target class as columns. For *extract class*, the lines are the fields and methods of the source class and the columns are the methods of the extracted class. For *move class*, the lines of the cohesion matrix are composed by the fields, methods of the class to move, and the columns are the fields and methods of the classes of the target package. Regarding *merge package* and *extract/split package* operations, the lines correspond to the fields and methods of the classes of the source package and the columns are the fields and methods of the classes of the target package. We calculate the similarity between each pair of elements in a cohesion matrix. The cohesion matrix is obtained as follows: For the field-field similarity, we consider the vocabulary-based similarity. For the method-method similarity, we consider both vocabulary and dependency-based similarity. For the method-field similarity, if the method $m_i$ may access (read or write) the field $f_j$, then the similarity value is 1. Otherwise, the similarity value is 0. For example, the suitable set of methods to be moved to a new class is obtained as follows: we consider the line with the highest average value and construct a set that consists of the elements in this line that have a similarity value that is higher than a certain threshold.

The semantic function for a remodularization operation corresponds to the average (equal importance) of the different three semantic measures described above. As a remodularization solution is a sequence of operations, the overall semantic evaluation of the solution is the average of the semantic values of all the operations composing a solution.

Some code changes contribute to the domain vocabulary of the system but not necessarily all of them. Thus, we are considering semantic coherence and consistency with prior code changes as separate objectives. In addition, the consistency with priori code changes is not mainly related to the domain vocabulary but to the type of operations that are applied to similar context. Treating the similarity with prior changes as a separate objective can address the problem that developers use sometimes names of code elements that semantically do not make any sense. Furthermore, we decided to separate semantic coherence and the consistency with prior code changes in two different objectives to give more flexibility to user when selecting the best solution based on their preferences and also to ensure the usefulness of our approach even in the situation when the change history is not available (the user can easily exclude this objective while maintaining the semantic coherence one).

To find a compromise between the seven objectives described in this section, we used a recent many-objective optimization algorithm (NSGA-III) that will be described in the next section.

## 4. SOFTWARE REMODULARIZATION USING NSGA-III

This section shows how the remodularization problem can be addressed using NSGA-III. We first present an overview of the technique then we provide the details of our adaptation to the remodularization problem.

### 4.1 NSGA-III

NSGA-III is a recent many-objective algorithm proposed by [Deb Jain 2014]. The basic framework remains similar to the original NSGA-II algorithm with significant changes in its selection mechanism. Figure 5 gives the pseudo-code of the NSGA-III procedure for a particular generation $t$. First, the parent population $P_t$ (of size $N$) is randomly initialized in the specified domain, and then the binary tournament selection, crossover and mutation operators are applied to create an offspring population $Q_t$. Thereafter, both populations are combined and sorted according to their domination level and the best $N$ members are selected from the combined population to form the parent population for the next generation. The fundamental difference between NSGA-II and NSGA-III lies in the way the niche preservation operation is performed. Unlike NSGA-II, NSGA-III starts with a set of reference points $Z^r$. The set of uniformly distributed reference points is generated using the method of [Das and Dennis 1998] which is well-detailed and described in [Asafuddoula et al. 2013].

In order to illustrate the reference point generation process, we give in what follows an example of such generation with only three objectives in order to ease the understanding and the visualization of such process. However, this process is generic for any number of objectives $M$. The Das and Dennis approach for this case generates $W$ reference points on the hyperplane with a uniform spacing $\delta = 1/p$. We assume for this example that $p = 3$ (i.e., $\delta = 0.33$). The number of reference points is thus $W = C_p^{(M+p-1)}$ which is equal to 10. The following figure describes the reference point set generation mechanism:

The following Figure 4 shows an Illustration of the Das and Dennis method for the generation of the reference points by computing $R1$, $R2$, and $R3$ recursively. The table shows the combinations of $R1$, $R2$, and $R3$ components. The figure shows the plotting of the 10 obtained reference points on the hyperplane. For our remodularization problem, we have seven objectives ($M = 7$) and we use $p = 5$. These are the only parameters to adjust for our remodularization problem.

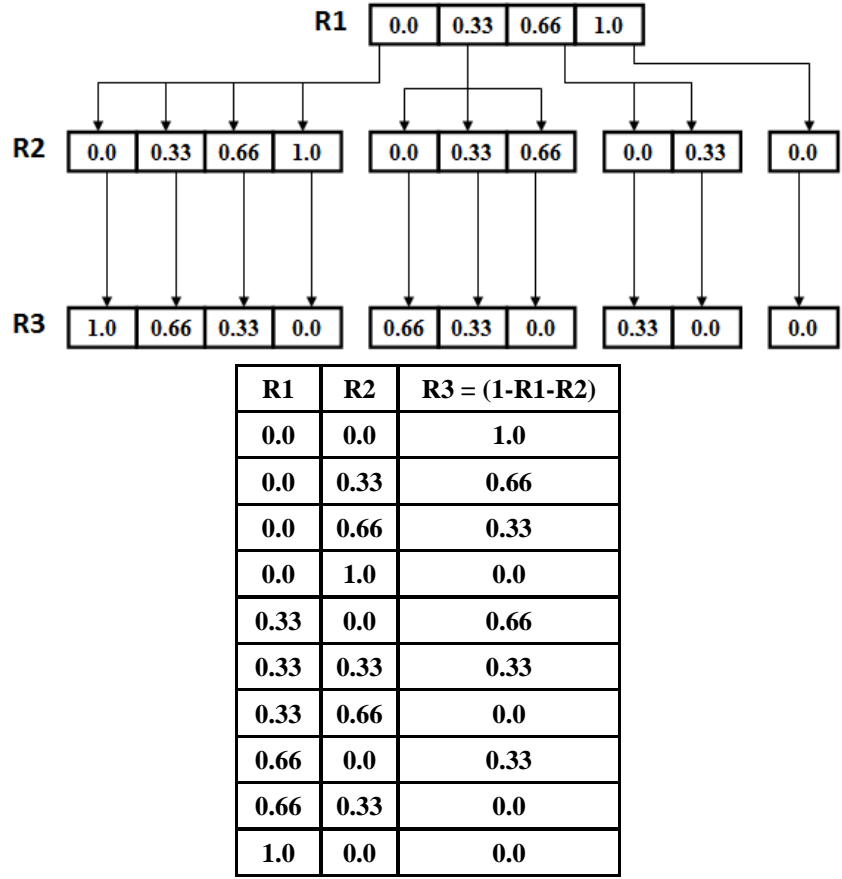

| R1 | R2 | R3 = (1-R1-R2) |
|---|---|---|
| 0.0 | 0.0 | 1.0 |
| 0.0 | 0.33 | 0.66 |
| 0.0 | 0.66 | 0.33 |
| 0.0 | 1.0 | 0.0 |
| 0.33 | 0.0 | 0.66 |
| 0.33 | 0.33 | 0.33 |
| 0.33 | 0.66 | 0.0 |
| 0.66 | 0.0 | 0.33 |
| 0.66 | 0.33 | 0.0 |
| 1.0 | 0.0 | 0.0 |

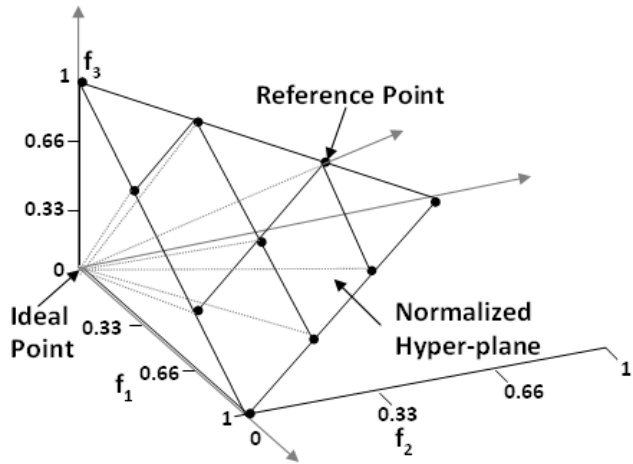

**Fig. 4.** 3D plotting of the obtained 10 reference points with p = 3.

After non-dominated sorting, all acceptable front members and the last front $F_l$ that could not be completely accepted are saved in a set $S_t$. Members in $S_t/F_l$ are selected right away for the next generation. However, the remaining members are selected from $F_l$ such that a desired diversity is maintained in the population. Original NSGA-II uses the crowding distance measure for selecting well-distributed set of points, however, in NSGA-III the supplied reference points ($Z^r$) are used to select these remaining members (cf. Figure 5). To accomplish this, objective values and reference points are first normalized so that they have an identical range. Thereafter, orthogonal distance between a member in $S_t$ and each of the reference lines (joining the ideal point, i.e., the vector composed of 7 zero and a reference point) is computed. The member is then associated with the reference point having the smallest orthogonal distance. Next, the niche count $\rho$ for each reference point, defined as the number of members in $S_t/F_l$ that are associated with the reference point, is computed for further processing. The reference point having the minimum niche count is identified and the member from the last front $F_l$ that is associated with it is included in the final population. The niche count of the identified reference point is increased by one and the procedure is repeated to fill up population $P_{t+1}$.

It is worth noting that a reference point may have one or more population members associated with it or need not have any population member associated with it. Let us denote this niche count as $\rho_j$ for the $j$-th reference point. We now devise a new niche-preserving operation as follows. First, we identify the reference point set $J_{\min} = \{j: \mathrm{argmin}_j (\rho_j)\}$ having minimum $\rho_j$. In case of multiple such reference points, one ($j^* \in J_{\min}$) is chosen at random. If $\rho_{j^*} = 0$ (meaning that there is no associated $P_{t+1}$ member to the reference point $j^*$), two scenarios can occur. First, there exists one or more members in front $F_l$ that are already associated with the reference point $j^*$. In this case, the one having the shortest perpendicular distance from the reference line is added to $P_{t+1}$. The count $\rho_{j^*}$ is then incremented by one. Second, the front $F_l$ does not have any member associated with the reference point $j^*$. In this case, the reference point is excluded from further consideration for the current generation. In the event of $\rho_{j^*} \geq 1$ (meaning that already one member associated with the reference point exists), a randomly chosen member, if exists, from front $F_l$ that is associated with the reference point $F_l$ is added to $P_{t+1}$. If such a member exists, the count $\rho_{j^*}$ is incremented by one. After $\rho_j$ counts are updated, the procedure is repeated for a total of $K$ times to increase the population size of $P_{t+1}$ to $N$.

### 4.2 Solution Approach

This section describes our adaptation of NSGA-III to our remodularization problem. Thus, we define the following adaptation steps: representation of the solutions and the generation of the initial population, evaluation of individuals using the fitness functions, selection of the individuals from one generation to another, generation of new individuals using genetic operators (crossover and mutation) to explore the search space and the normalization of population members.

*4.2.1 Solution Representation*

To represent a candidate remodularization solution (individual), we used a vector representation. Each vector's dimension represents a remodularization operation. Thus, a solution is defined as a sequence of operations applied to different parts of the system to improve its modularization. A randomly generated solution is created as follows. First, we generate the solution length randomly between the lower and upper bounds of the solution length. After that, for each chromosome dimension, we generate a number $i$ between 1 and the total number of possible operations, then we assign the $i^{th}$ operation to the considered dimension. For each operation, the parameters, described in Table 2, are randomly generated from the list of source code elements extracted from the system to remodularize using a parser. An example of a solution is given in Figure 6 applied to the motivating example described in Section 2.

When generating a sequence of operations (individual), it is important to guarantee that they are feasible and that they can be applied. The first work in the literature was proposed by [Opdyke 1992] who introduced a way of formalizing the preconditions that must be imposed before a code change can be applied in order to preserve the behavior of the system. Opdyke created functions which could be used to formalize constraints. These constraints are similar to the Analysis Functions used later by [O'Keeffe and Ó Cinnéide 2008; Ó Cinnéide 0et al. 2012] who developed a tool to reduce program analysis. In our approach, we used a system to check a set of simple conditions, taking inspiration from the work proposed by Ó Cinnéide. Our search-based remodularization tool simulates operations using pre and post conditions that are expressed in terms of conditions on a code model. For example, to apply the remodularization operation *MoveClass(GanttTaskRelationship, net.sourceforge.ganttproject, net.sourceforge.ganttproject.task)*, a number of necessary preconditions should be satisfied, e.g.,

*net.sourceforge.ganttproject* and *net.sourceforge.ganttproject.task* should exist and should be packages; *GanttTaskRelationship* should exist and should be a class and the class *AttrNSImpl* should be implemented in the package *net.sourceforge.ganttproject*. As postconditions, *GanttTaskRelationship, net.sourceforge.ganttproject,* and *net.sourceforge.ganttproject.task* should exist; *GanttTaskRelationship* class should be in the package *net.sourceforge.ganttproject.task* and should not exists anymore in the package *net.sourceforge.ganttproject*.

**NSGA-III procedure at generation $t$**

**Input:** $H$ structured reference points $Z^s$, parent population $P_t$
**Output:** $P_{t+1}$
00: **Begin**
01: $S_t \leftarrow \emptyset, i \leftarrow 1$;
02: $Q_t \leftarrow$ Variation $(P_t)$;
03: $R_t \leftarrow P_t \cup Q_t$;
04: $(F_1, F_2, ...) \leftarrow$ Non-dominationed_Sort $(R_t)$;
05: **Repeat**
06:    $S_t \leftarrow S_t \cup F_i; i \leftarrow i+1$;
07: **Until** $| S_t | \geq N$;
08: $F_l \leftarrow F_i$; /*Last front to be included*/
09: **If** $| S_t | = N$ **then**
10:    $P_{t+1} \leftarrow S_t$;
11: **Else**
12:    $P_{t+1} \leftarrow \bigcup_{j=1}^{l-1} F_j$;
   /*Number of points to be chosen from $F_l$*/
13:    $K \leftarrow N - |P_{t+1}|$;
   /*Normalize objectives and create reference set $Z^r$*/
14:    Normalize $(F^M; S_t; Z^r; Z^s)$;
   /*Associate each member $s$ of $S_t$ with a reference point*/
   /*$\pi(s)$: closest reference point*/
   /*$d(s)$: distance between $s$ and $\pi(s)$*/
15:    $[\pi(s), d(s)] \leftarrow$ Associate $(S_t, Z^r)$;
   /*Compute niche count of reference point $j \in Z^r$ */
16:    $\rho_j \leftarrow \sum_{s \in S_t / F_l} ((\pi(s) = j) ? 1 : 0)$;
   /*Choose $K$ members one at a time from $F_l$ to construct $P_{t+1}$*/
17:    Niching $(K, \rho_j, \pi(s), d(s), Z^r, F_l, P_{t+1})$;
18: **End If**
19: **End**

**Input:** *Sys*: system to evaluate, $P_t$: parent population, $P_u$: Updated population
**Output:** $P_{t+1}$

```
00:  Begin
01:  If UserFeedback = TRUE then
02:      P_t ← User_Feedback (P_u);
03:      ...Sys ← Refactored_system ();
04:      UserFeedback ← FALSE;
05:  End If
06:  S_t ← Ø, i ← 1;
07:  Q_t ← Variation (P_t);
08:  R_t ← P_t ∪ Q_t;
09:  P_t ← evaluate (P_t, Sys);
10:  (F_1, F_2, ...) ← Non-dominationed_Sort (R_t);
11:  Repeat
12:      S_t ← S_t ∪ F_i; i ← i+1;
     Until | S_t | ≥ N;
13:  F_l ← F_i; //Last front to be included*/
     If | S_t | = N then
14:      P_{t+1} ← S_t;
15:  Else
         P_{t+1} ← ⋃_{j=1}^{l-1} F_j;
15:      /*Number of points to be chosen from F_l*/
         K ← N – |P_{t+1}|;
         /*Crowding distance of points in F_l */
15:      Crowding-Distance-Assignment(F_l);
         ...Sort(F_l);
         /*Choose K solutions with largest distance*/
16:      P_{t+1} ← P_{t+1} ∪ Select(F_l, k);
     End If
17:  If t+1 = Threshold then
18:      UserFeedback ← TRUE;
19:      ...P_t ← User_Feedback (P_u);
     End If
     End
```

**Fig. 5.** Pseudo-code of NSGA-III main procedure.

| |
|---|
| Move Class(GanttTaskRelationship, net.sourceforge.ganttproject, net.sourceforge.ganttproject.task) |
| Extract Class(XGrammarWriter, XGrammarInput, parseInt()) |
| Move Method(normalize(), XGrammarWriter, DTDGrammar) |
| Extract Package(net.sourceforge.ganttproject, net.sourceforge.ganttproject.dtl, CharacterDataImpl, ChildNode) |

**Fig. 6.** NSGA-III solution representation.

*4.2.2 Fitness Functions*

Each generated remodularization solution is executed on the system *S*. Once all required data is computed, the solution is evaluated based on the 7 objectives described in Section 3. Based on these values, the remodularization solution is assigned a non-domination rank (as in NSGA-II) and a position in the objective space allowing it to be assigned to a particular reference point based on distance calculation as previously described. As a reminder, the following fitness functions are used: 1) *number of classes per package* (to minimize); 2) *number of packages in the system* (to minimize); 3) *cohesion* (to maximize); 4) *coupling* (to minimize); 5) *Semantics coherence* (to maximize); 6) *number of operations* (to minimize); and 7) *coherence with the history of code changes* (to minimize). The *semantic fitness function* of a solution corresponds to the average of the semantic values of the operations in the vector. The *history of changes fitness function* maximizes the use of ROs that are similar to those applied to the same code fragments in the past. To calculate the similarity score between a proposed remodularization operation and a recorded operation, we use the fitness function described in Section 3.

**Normalization of population members.** Usually objective functions are incommensurable (i.e., they have different scales). For this reason, we used the normalization procedure proposed by [0Deb and Jain 2014] to circumvent this problem. At each generation, the minimal and maximal values for each metric are recorded and then used by the normalization procedure. Normalization allow the population members and with the reference points to have the same range, which is a pre-requisite for diversity preservation.

*4.2.3    Evolutionary Operators*

In each search algorithm, the variation operators play the key role of moving within the search space with the aim of driving the search towards optimal solutions.

For crossover, we use the one-point crossover operator. It starts by selecting and splitting at random two parent solutions. Then, this operator creates two child solutions by putting, for the first child, the first part of the first parent with the second part of the second parent, and vice versa for the second child. This operator must ensure the respect of the length limits by eliminating randomly some ROs. As illustrated in Figure 7, each child combines some of the operations of the first parent with some ones of the second parent. In any given generation, each solution will be the parent in at most one crossover operation. It is important to note that in many-objective optimization, it is better to create children that are close to their parents in order have a more efficient search process. For this reason, we control the cutting point of the one-point crossover operator by restricting its position to be either belonging to the first third of the operations sequence or belonging to the last third.

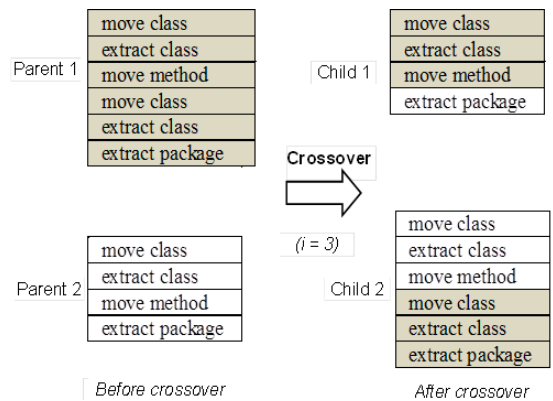

**Fig. 7.** Crossover operator.

For mutation, we use the bit-string mutation operator that picks probabilistically one or more operations from its or their associated sequence and replace them by other ones from the initial list of possible operations as described in the running example of Figure 8. The number of changes is unknown a priori and depends on the mutation probability. Indeed, each chromosome dimension would be changed according to the mutation probability. For example, for a mutation probability of 0.2, for each dimension, we generate randomly a number x between 0 and 1, if x<0.2 we change the dimension, otherwise not.

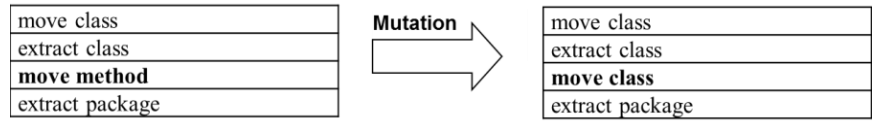

**Fig. 8.** Mutation operator.

After applying genetic operators (mutation and crossover), we verify the feasibility of the generated sequence of operations by checking the pre and post conditions. Each operation that is not feasible due to unsatisfied preconditions will be removed from the generated remodularization sequence. The new sequence is considered valid in our NSGA-III adaptation if the number of rejected operations is less than 10% of the total sequence size.

## 5. VALIDATION

In order to evaluate our approach for restructuring systems using NSGA-III, we conducted a set of experiments based on different versions of large open source systems and one industrial project provided by Ford Motor Company. Each experiment is repeated 31 times, and the obtained results are subsequently statistically analyzed with the aim to compare our NSGA-III proposal with a variety of existing approaches [Abdeen et al. 2009; Abdeen et al. 2013; Bavota et al. 2013]. In this section, we first present our research questions and then describe and discuss the obtained results. Finally, we discuss the various threats to the validity of our experiments.

### 5.1 Research Questions

In our study, we assess the performance of our remodularization approach by finding out whether it could generate meaningful sequences of operations that improve the structure of packages while reducing the number of code changes, preserving the semantic coherence of the design, and reusing as much as possible a base of recorded operations applied in the past in similar contexts. Our study aims at addressing the following research questions outlined below. We also explain how our experiments are designed to address these questions. The main question to answer is to what extent the proposed approach can propose meaningful remodularization solutions. To find an answer, we defined the following 7 research questions:

**RQ1.1**: To what extent can the proposed approach improve the structure of packages in the system?
**RQ1.2**: To what extent the proposed approach preserve the semantics while improving the packages structure?
**RQ1.3**: To what extent can the proposed approach minimize the number of changes (size)?
**RQ1.4**: To what extent the use of recorded changes improve the suggestion of good remodularization solutions?
**RQ2**: How does the proposed many-objective approach based on NSGA-III perform compared to other many/multi-objective algorithms or a mono-objective approach?
**RQ3**: How does the proposed many-objective approach based on NSGA-III perform compared to existing remodularization approach not based on heuristic search?
**RQ4:** Insight. How our many-objective remodularization approach can be useful for software engineers in real-world setting?

To answer RQ1.1, we validate the proposed remodularization on four medium to large-size open-source systems and one industrial project to evaluate the structural improvements of systems after applying the best solution. To this end, we used the following metrics: *average number of classes per package (NCP), number of packages (NP), number of inter-edges (NIE) and number of intra-edges (NAE)*.

To answer RQ1.2, it is important to validate the proposed remodularization solutions from both quantitative and qualitative perspectives. To this end, we use two different validation methods: manual validation and automatic validation of the efficiency of the proposed solutions. For the manual validation, we asked groups of potential users (software engineers) of our remodularization tool to evaluate, manually, whether the suggested operations are feasible and make sense semantically. We define the metric "manual precision" (MP) which corresponds to the number of meaningful operations, in terms of semantic coherence, over the total number of suggested operations. MP is given by the following equation

$$MP = \frac{|coherent\ operations|}{|proposed\ operations|} \in [0,1] \quad \quad (8)$$

For the automatic validation, we introduce manually several changes on the remodularization of JHotDraw and we evaluate the ability of our approach to generate the initial version of the system (considered as a well-designed system). In fact, JHotDraw is considered as one of the well-designed open source systems and several design patterns are used in its implementation. Thus, we compare the proposed operations with the expected ones in terms of recall and precision:

$$RE_{recall} = \frac{|suggested\ operations| \cap |expected\ operations|}{|expected\ operations|} \in [0,1] \qquad (9)$$

$$PR_{precesion} = \frac{|suggested\ operations| \cap |expected\ operations|}{|suggested\ operations|} \in [0,1] \qquad (10)$$

To answer RQ1.3, we evaluate the number of operations (*NO*) suggested by the best remodularization solutions on the different systems.

To answer RQ1.4, we use the metric *MP* to evaluate the effect of the use of recorded operations, applied in the past to similar contexts, on the semantic coherence. Moreover, in order to evaluate the importance of reusing recorded operations in similar contexts, we define the metric "reused operations" (ROP) that calculates the percentage of operations from the base of recorded operations used to generate the optimal remodularization solutions by our proposal. *ROP* is given by the following equation

$$ROP = \frac{|used\ operations\ from\ the\ base\ of\ recorded\ operations|}{|base\ of\ recorded\ operations|} \in [0,1] \qquad (11)$$

To answer RQ2, we compared the performance of NSGA-III with two many-objective techniques, MOEA/D and IBEA, and also with a multi-objective algorithm that uses NSGA-II. We used *Inverted Generational Distance (IGD)* to compare between the different algorithms*:* A number of performance metrics for multi-objective optimization have been proposed and discussed in the literature, which aim to evaluate the closeness to the Pareto optimal front and the diversity of the obtained solution set, or both criterion. Most of the existing metrics require the obtained set to be compared against a specified set of Pareto optimal reference solutions. In this study, the inverted generational distance (IGD) is used as the performance metric since it has been shown to reflect both the diversity and convergence of the obtained non-dominated solutions. The IGD corresponds to the average Euclidean distance separating each reference solution from its closest non-dominated one. Note that for each system we use the set of Pareto optimal solutions generated by all algorithms over all runs as reference solutions. In addition to *IGD*, we used the above described metrics to compare between all the algorithms: *NCP, NP, NIE, NAE, MP, RE,* and *PR*. We also compared our approach with a multi-objective remodularization technique proposed by [Abdeen et al. 2013] where the objectives considered are coupling, cohesion and number of changes. Since the approach of Abdeen et al. is limited to the use of only one operation (Move Class), we only used the qualitative evaluation based on *MP* for the comparison.

It is important also to determine if considering each conflicting metric as a separate objective to optimize performs better than a mono-objective approach that aggregates several metrics in one objective. The comparison between a many-objective EA with a mono-objective one is not straightforward. The first one returns a set of non-dominated solutions while the second one returns a single optimal solution. In order to resolve this problem, for each many-objective algorithm we choose the nearest solution to the Knee point [Bechikh et al. 2011] (i.e., the vector composed of the best objective values among the population members) as a candidate solution to be compared with the single solution returned by the mono-objective algorithm. We compared NSGA-III with an existing mono-objective remodularization approach [Abdeen et al. 2009] based on the use of cohesion and coupling aggregated in one fitness function. Since the mono-objective approach is limited to the use of only one operation (Move Class), we only used the qualitative evaluation based on *MP* for the comparison and feed-back from software engineers on using both tools.

For RQ3, since it is not sufficient to outperform existing search-based remodularization techniques, we compared our proposal to an existing remodularization technique based on the use of coupling and cohesion [Bavota et al. 2013] and limited to the only use of "Split packages" change. Thus, we compared our proposal using only the qualitative evaluation based on *MP* and feed-back from software engineers on using both tools.

For RQ4, we evaluated the benefits of our remodularization tool by several software engineers. To this end, they classify the suggested operations (*IOP*) one by one as interesting or not. The difference with the *MP* metric is that the operations are not classified from a semantic coherence perspective but form a usefulness one.

$$IOP = \frac{|useful operations|}{|operations|} \in [0,1] \qquad (12)$$

To answer the above research questions, we selected the solution from the set of non-dominated ones providing the maximum trade-off using the following strategy when comparing between the different algorithms (expect the mono-objective algorithm where we select the solution with the highest fitness function). In order to find the maximal trade-off solution of the multi-objective or many-objective algorithm, we use the trade-off worthiness metric proposed by [Rachmawati and Srinivasan 2009] to evaluate the worthiness of each non-dominated solution in terms of compromise between the objectives. This metric is expressed as follows:

$$\mu(x_i, S) = \underset{x_j \in S, x_i \not\prec x_j, x_j \not\prec x_i}{Min} T(x_i, x_j) \qquad (13)$$

where
$$T(x_i, x_j) = \frac{\sum_{m=1}^{M} \max\left[0, \frac{f_m(x_j) - f_m(x_i)}{f_m^{\max} - f_m^{\min}}\right]}{\sum_{m=1}^{M} \max\left[0, \frac{f_m(x_i) - f_m(x_j)}{f_m^{\max} - f_m^{\min}}\right]}$$

We note that $x_j$ denotes members of the set of non-dominated solutions $S$ that are non-dominated with respect to $x_i$. The quantity $\mu(x_i, S)$ expresses the least amount of improvement per unit deterioration by substituting any alternative $x_j$ from $S$ with $x_i$. We note also that $f_m(x_i)$ corresponds to the $m^{th}$ objective value of solution $x_i$ and $f_m^{\max}/f_m^{\min}$ corresponds to the maximal/minimal value of the $m^{th}$ objective in the population individuals. In the above equations, normalization is performed in order to prevent some objectives being predominant over others since objectives are usually incommensurable in real world applications. In the last equation, the numerator expresses the aggregated improvement gained by substituting $x_j$ with $x_i$. However, the denominator evaluates the deterioration generated by the substitution.

### 5.2 Software Projects Studied

We used a set of well-known open-source java projects and one project from our industrial partner Ford Motor Company. We applied our approach to four large and medium size open-source java projects: Xerces-J, JFreeChart, GanttProject, and JHotDraw. Xerces-J is a family of software packages for parsing XML. JFreeChart is a powerful and flexible Java library for generating charts. GanttProject is a cross-platform tool for project scheduling. JHotDraw is a GUI framework for drawing editors. Finally, the industrial project, JDI, is Java-based software system that helps Ford Motor Company analyze useful information from the past sales of dealerships data and suggests which vehicles to order for their dealer inventories in the future. This system is main key software application used by Ford Motor Company to improve their vehicles sales by selecting the right vehicle configuration to the expectations of customers. JDI is a highly structured and several versions were proposed by software engineers at Ford during the past 10 years. Due to the importance of the application and the high number of updates performed during a period of 10 years, it is critical to ensure good modularization of JDI to reduce the time required by developers to introduce new features in the future.

We selected these systems for our validation because they range from medium to large-sized open-source projects, which have been actively developed over the past 10 years. Table 4 provides some descriptive statistics about these six programs.

As described in Table 4, the upper and lower bounds on the chromosome length used in this study are set to 10 and 350 respectively. Several SBSE problems including remodularization are characterized by a varying chromosome length. This issue is similar to the problem of bloat control in genetic programming where the goal is to identify the tree size limits. To solve this problem, we performed several trial and error experiments where we assess the average performance of NSGA-III using the HV (hypervolume) performance indicator while varying the size limits between 10 and 500 operations. Figure 9 shows the obtained results which explains our choices described in Table 4.

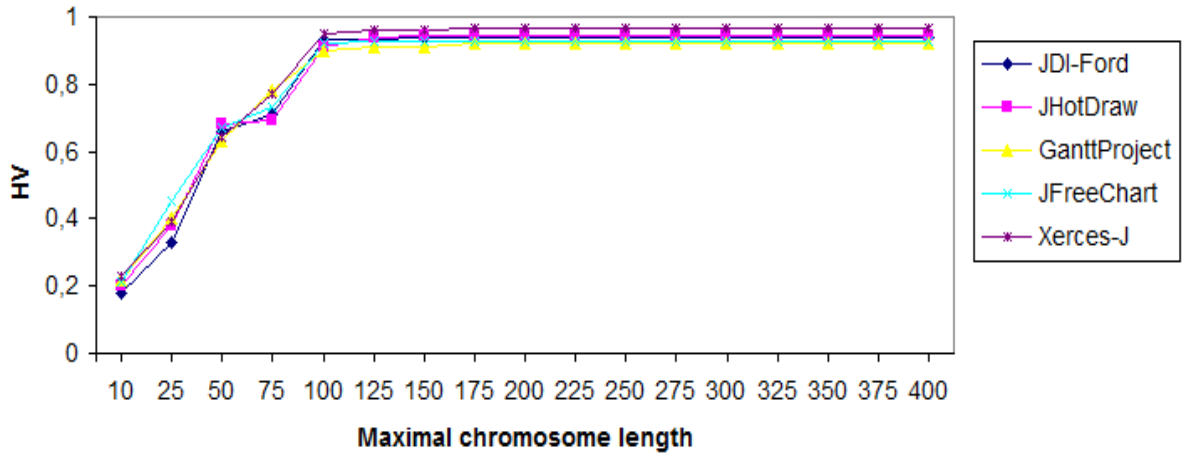

**Fig. 9.** Average Performance of NSGA-III the search algorithms using the HV (hypervolume) indicator while varying the size limits between 10 and 500 operations

**Table 4.** Statistics of the studied systems and solution length limits.

| Systems | Release | # classes | KLOC | Solution length limits (Min/Max) |
|---|---|---|---|---|
| Xerces-J | v2.7.0 | 991 | 240 | [35 , 350] |
| JHotDraw | v6.1 | 585 | 21 | [20 , 250] |
| JFreeChart | v1.0.9 | 521 | 170 | [20 , 250] |
| GanttProject | v1.10.2 | 245 | 41 | [10 , 150] |
| JDI-Ford | v5.8 | 638 | 247 | [25 , 300] |

To collect operations applied in previous program versions, we use Ref-Finder [Prete et al. 2010]. Ref-Finder, implemented as an Eclipse plug-in, can identify operations between two releases of a software system. Table 5 shows the analyzed versions and the number of operations, identified by Ref-Finder, between each subsequent couple of analyzed versions, after the manual validation. In our study, we consider only the operation types described in Table 2.

**Table 5.** Analyzed versions and operations collection

| Systems | Collected operation | |
|---|---|---|
| | Previous releases | # operations |
| Xerces-J | v1.4.2 - v2.6.1 | 52 |
| JFreeChart | v1.0.6 - v1.0.8 | 63 |
| GanttProject | v1.7 - v1.10.1 | 81 |
| JHotDraw | v5.1 - v6.0 | 56 |
| JDI-Ford | v2.4 – v5.6 | 97 |

### 5.3 Experimental Setting

The goal of the study is to evaluate the usefulness and the effectiveness of our remodularization tool in practice. We conducted a non-subjective evaluation with potential developers who can use our tool. Indeed, operations should not only improve the structure of packages, but should also be meaningful from a developer's point of view in terms of semantic coherence and usefulness.

*5.3.1 Subjects*

Our study involved 13 subjects from the University of Michigan and 2 software engineers from Ford Motor Company. Subjects include 5 master students in Software Engineering, 7 PhD students in Software Engineering, 1 faculty member in Software Engineering, and 2 junior software developers. 4 of them are females and 11 are males. All the subjects are volunteers and familiar with Java development. The experience of these subjects on Java programming ranged from 2 to 16 years. The evaluated solutions by the subjects are those that represent the maximum trade-off between the objective using the trade-off worthiness metric proposed by Rachmawati and Srinivasan as described in Section 5.1.

*5.3.2   Scenario*

We designed our study to answer our research questions. The subjects were invited to fill a questionnaire that aims to evaluate our suggested operations. We divided the subjects into six groups according to 1) the number of studied systems (Table 5) the number of remodularization solutions to evaluate, and 3) the number of techniques to be tested.

The number of remodularization solutions to evaluate depends on different objectives combinations: *average number of classes per package (NCP), number of packages (NP), coupling (COU), cohesion (COH), semantics preservation (SP), number of changes (NCH) and coherence with history of changes (CHC)*. For each combination (two, three, four, five, six and seven, objectives), a remodularization solution is suggested to find the best compromise between the considered objectives. Similarly, the solutions of the state-of-the art works are empirically evaluated in order to compare them to our approach as described in the previous section. Table 6 describes the number of remodularization solutions to be evaluated for each studied system in order to answer our research questions.

**Table 6.** Considered solutions for the qualitative evaluation

| Ref. Solution | Algorithm/ Approach | # Objectives | Considered objectives |
|---|---|---|---|
| Solution 1 | NSGA-III | 2 | NCP, NP |
| Solution 2 | | 3 | NCP, NP, COU |
| Solution 3 | | 4 | NCP, NP, COU, COH |
| Solution 4 | | 5 | NCP, NP, COU, COH, SP |
| Solution 5 | | 6 | NCP, NP, COU, COH, SP, NCH |
| Solution 6 | | 7 | NCP, NP, COU, COH, SP, NCH, CHC |
| Solution 7 | IBEA | 2 | NCP, NP |
| Solution 8 | | 3 | NCP, NP, COU |
| Solution 9 | | 4 | NCP, NP, COU, COH |
| Solution 10 | | 5 | NCP, NP, COU, COH, SP |
| Solution 11 | | 6 | NCP, NP, COU, COH, SP, NCH |
| Solution 12 | | 7 | NCP, NP, COU, COH, SP, NCH, CHC |
| Solution 13 | MOEA/D | 2 | NCP, NP |
| Solution 14 | | 3 | NCP, NP, COU |
| Solution 15 | | 4 | NCP, NP, COU, COH |
| Solution 16 | | 5 | NCP, NP, COU, COH, SP |
| Solution 17 | | 6 | NCP, NP, COU, COH, SP, NCH |
| Solution 18 | | 7 | NCP, NP, COU, COH, SP, NCH, CHC |
| Solution 19 | Mono-objective Simulate Annealing [Abdeen et al. 2011] | 1 | COU+COH |
| Solution 20 | NSGA-II [Abdeen et al. 2013] | 3 | COU, COH, NCH |
| Solution 21 | Automated re-modularization [Bavota et al. 2013] | 2 | COU, COH |

As shown in Table 6, for each system, 21 remodularization solutions have to be evaluated. Due to the huge number of operations to be evaluated (each solution consists of a set of operations), we pick at random a sub-set of up-to 10 operations per solution to be evaluated in our study. In Table 7, we summarize how we divided subjects into 6 groups in order to cover all remodularization solutions. In addition, as illustrated in Table 7, we are using a cross-validation to reduce the impact of subjects on the evaluation. Each subject evaluates different remodularization solutions for three different systems.

Subjects were first asked to fill out a pre-study questionnaire containing five questions. The questionnaire helped to collect background information such as their role within the company, their programming experience, their familiarity with software remodularization. In addition, all the participant attended one lecture about software remodularization and passed five tests to evaluate their performance to evaluate and suggest remodularization solutions. Then, the groups are formed based on the pre-study questionnaire and the tests result to make sure that all the groups have almost the same average skills.

The participants were asked to justify their evaluation of the solutions and these justifications are reviewed by the organizers of the study (one faculty member, one postdoc, one PhD student and one master student). In addition, our experiments are not only limited to the manual validation but also the automatic validation can verify the effectiveness of our approach.

Subjects were aware that they are going to evaluate the semantic coherence and the usefulness of the operations, but do not know the particular experiment research questions (algorithms used, different objectives used and their combinations). Consequently, each group of subjects who accepted to participate to the study, received a questionnaire, a manuscript guide to help them to fill the questionnaire, and the source code of the studied systems, in order to evaluate 21 solutions (10 operations per solution). The questionnaire is organized in an excel file with hyperlinks to visualize easily the source code of the affected code elements. Subjects are invited to select for each operation one of the possibilities: *"Yes"*, *"No"*, or *"May be"* (if not sure) about the semantic coherence and usefulness. Since the application of remodularization solutions is a subjective process, it is normal that not all the programmers have the same opinion. In our case, we considered the majority of votes to determine if suggested solutions are correct or not.

Table 7. Survey organization

| Subject groups | Systems | Algorithms / Approaches | Solutions |
| --- | --- | --- | --- |
| Group A | GanttProject | NSGA-III<br>IBEA | Solution 1-6<br>Solution 7-12 |
|  | Xerces | MOEA/D,<br>Abdeen et al.2011 | Solution 13-18<br>Solution 19 |
|  | JFreeChart | Abdeen et al. 2013,<br>Bavota et al. 2013 | Solution 20<br>Solution 21 |
| Group B | GanttProject | NSGA-III<br>IBEA | Solution 1-6<br>Solution 7-12 |
|  | Xerces | MOEA/D,<br>Abdeen et al.2011 | Solution 13-18<br>Solution 19 |
|  | JFreeChart | Abdeen et al. 2013,<br>Bavota et al. 2013 | Solution 20<br>Solution 21 |
| Group C | GanttProject | NSGA-III<br>IBEA | Solution 1-6<br>Solution 7-12 |
|  | Xerces | MOEA/D,<br>Abdeen et al.2011 | Solution 13-18<br>Solution 19 |
|  | JFreeChart | Abdeen et al. 2013,<br>Bavota et al. 2013 | Solution 20<br>Solution 21 |
| Group D | GanttProject | NSGA-III<br>IBEA | Solution 1-6<br>Solution 7-12 |
|  | JHotDraw | MOEA/D,<br>Abdeen et al.2011 | Solution 13-18<br>Solution 19 |
|  | JDI-Ford | Abdeen et al. 2013,<br>Bavota et al. 2013 | Solution 20<br>Solution 21 |
| Group E | Xerces | NSGA-III<br>IBEA | Solution 1-6<br>Solution 7-12 |
|  | JHotDraw | MOEA/D,<br>Abdeen et al.2011 | Solution 13-18<br>Solution 19 |
|  | JDI-Ford | Abdeen et al. 2013,<br>Bavota et al. 2013 | Solution 20<br>Solution 21 |
| Group F | JFreeChart | NSGA-III<br>IBEA | Solution 1-6<br>Solution 7-12 |
|  | JHotDraw | MOEA/D,<br>Abdeen et al.2011 | Solution 13-18<br>Solution 19 |
|  | JDI-Ford | Abdeen et al. 2013,<br>Bavota et al. 2013 | Solution 20<br>Solution 21 |

*5.3.3  Parameters Tuning*

Parameter setting influences significantly the performance of a search algorithm on a particular problem [Arcuri and Frazer 2013]. For this reason, for search-based algorithm and for each system (cf. Table 8), we perform a set of experiments using several population sizes: 62, 100, 150, 180, 140, and 190 for respectively 2, 3, 4, 5, 6 and 7 objectives.

The maximum number of generations used is 300, 500, 700, 1000, 1200, and 1400 respectively for 2, 3, 4, 5, 6 and 7 objectives. For each algorithm, to generate an initial population, we start by defining the maximum vector length (maximum number of operations per solution). The vector length is proportional to the number of operations that are considered and the size of the program to be restructured (cf. Table 5). A higher number of operations in a solution do not necessarily mean that the results will be better. Ideally, a small number of operations should be sufficient to provide a good trade-off between the fitness functions. This parameter can be specified by the user or derived randomly from the sizes of the program and the used operations list. During the creation, the solutions have random sizes inside the allowed range. Each algorithm is executed 31 times with each configuration and then comparison between the configurations is done based on IGD using the Wilcoxon test. In order to have significant results, for each couple (algorithm, system), we use the trial and error method 0 in order to obtain a good parameter configuration. Since we are comparing different search algorithms, we classify parameters into common parameters and specific parameters.

Table 8 depicts the important common parameters. For MOEA/D, the neighborhood size is set to 20. For IBEA, the scaling parameter κ is set to 0.01. For the SA of [Abdeen et al. 2011], the start and stop temperatures are set respectively to 22.8 and 1.0 using a geometrical cooling scheme of with a cooling rate of 0.9975 and the number of local search iterations is set to 15. It is important to note that all heuristic algorithms have the same termination criterion for each experiment (same number of evaluations) in order to ensure fairness of comparisons.

**Table 8.** The setting of common parameters.

| Number of objectives | Number of reference points (for NSGA-III and MOEA/D) | Population size | Number of generations | Crossover rate | Mutation rate |
|---|---|---|---|---|---|
| 2 | 62 | 100 | 300 | 0.9 | 0.1 |
| 3 | 100 | 150 | 500 | 0.9 | 0.1 |
| 4 | 150 | 182 | 700 | 0.9 | 0.1 |
| 5 | 180 | 174 | 1000 | 0.8 | 0.2 |
| 6 | 140 | 186 | 1200 | 0.8 | 0.2 |
| 7 | 190 | 193 | 1400 | 0.8 | 0.2 |

#### 5.4 Statistical Tests

Since metaheuristic algorithms are stochastic optimizers, they can provide different results for the same problem instance from one run to another. For this reason, our experimental study is performed based on 31 independent simulation runs for each problem instance and the obtained results are statistically analyzed by using the Wilcoxon rank sum test [Arcuri and Briand 2011] with a 99% confidence level (α = 1%). The latter verifies the null hypothesis H0 that the obtained results of two algorithms are samples from continuous distributions with equal medians, against the alternative that they are not H1. The p-value of the Wilcoxon test corresponds to the probability of rejecting the null hypothesis H0 while it is true (type I error). A p-value that is less than or equal to α (≤ 0.01) means that we accept H1 and we reject H0. However, a p-value that is strictly greater than α (> 0.01) means the opposite. In fact, for each problem instance, we compute the p-value obtained by comparing NSGA-II, IBEA, MOEA/D and mono-objective search results with NSGA-III ones. In this way, we determine whether the performance difference between NSGA-III and one of the other approaches is statistically significant or just a random result.

#### 5.5 Results

*5.5.1 Results for RQ1.1*

Table 9 summarizes the results of median values of the structural metrics over 31 independent simulation runs after applying the proposed operations by the remodularization solution selected using the knee-point strategy. The results of Table 9 are based on the consideration of all the 7 objectives for the many-objective algorithms thus the order is not important and have no impact on the results. The software engineer can select the best solution based on his preferences (fitness function values) and programming behavior from the non-dominated (trade-off) set of solutions.

As described in Table 9, we found that NSGA-III algorithm provides similar structural improvements the other techniques in terms of average number of classes per package (NCP), cohesion (NIE) and coupling (NAE). However, the number of packages (NP) in the system after applying NSGA-III solutions is slightly higher than all NP values proposed by the best solutions of most of the remaining algorithms in most of the cases except Xerces-J. This can be explained by the fact that decreasing the number of classes per package will automatically increase the number of packages. In addition, the extract class operation created new classes that increased the average number of classes per package.

The structural improvement scores of multi-objective and mono-objective algorithms are very close to those produced by many-objective algorithms especially NSGA-III. This is an interesting result confirming that our NSGA-III can find very good compromises between 7 objectives that are similar and sometimes outperforms those that are produced by existing approaches using only structural and semantic objectives.

We believe that improving the structure of packages it is a difficult and very important objective to reach. We consider that NSGA-III performance in terms of improving the structure similar to existing approaches is a very interesting result since the main goal of this work is to improve the structure while preserving the domain semantics which not well-considered by the remaining approaches.

**Table 9.** Average number of classes per package (NCP), number of packages (NP), number of inter-edges (NIE), number of intra-edges (NAE) and the deviation (delta with the initial design) median values of NSGA-III, IBEA, MOEA/D, SA Abdeen et al. 2011, NSGA-II Abdeen et al.2011 and Bavota et al. 2013 over 31 independent simulation runs. A "+" symbol at the ith position in the sequence of signs presented below the instance names means that the NSGA-III algorithm metric median value is statistically different from the ith algorithm one. A "-" symbol at the ith position in the sequence of signs means the opposite (e.g., for Xerces-J, NSGA-III is not statistically different from IBEA, however, it is statistically different from the other algorithms).

| System | Approach | NCP | dev.NCP | NP | dev.NP | NIE | dev.NIE | NAE | Dev.NAE |
|---|---|---|---|---|---|---|---|---|---|
| Xerces-J (-++++) | NSGA-III | 19 | -4 | 46 | +5 | 316 | -69 | 432 | +72 |
| | IBEA | 21 | -2 | 44 | +3 | 328 | -57 | 411 | +51 |
| | MOEA/D | 18 | -5 | 56 | +15 | 352 | -33 | 397 | +37 |
| | SA Abdeen et al. 2011 | 24 | +1 | 42 | +1 | 314 | -71 | 441 | +81 |
| | NSGA-II Abdeen et al.2011 | 18 | -5 | 56 | +15 | 333 | -52 | 422 | +62 |
| | Bavota et al. 2013 | 18 | -5 | 56 | +15 | 302 | -83 | 453 | +93 |
| JFreeChart (+++++) | NSGA-III | 14 | -6 | 38 | +8 | 286 | -71 | 384 | +69 |
| | IBEA | 16 | -4 | 36 | +6 | 304 | -53 | 392 | +77 |
| | MOEA/D | 16 | -4 | 36 | +6 | 314 | -43 | 383 | +86 |
| | SA Abdeen et al. 2011 | 13 | -7 | 42 | +12 | 291 | -66 | 396 | +81 |
| | NSGA-II Abdeen et al.2011 | 13 | -7 | 42 | +12 | 301 | -56 | 3786 | +71 |
| | Bavota et al. 2013 | 11 | -9 | 47 | +17 | 278 | -79 | 398 | +94 |
| GanttProject (--+++) | NSGA-III | 14 | -3 | 18 | +9 | 259 | -68 | 294 | +81 |
| | IBEA | 12 | -5 | 21 | +12 | 247 | -80 | 304 | +91 |
| | MOEA/D | 14 | -3 | 18 | +9 | 259 | -68 | 291 | +78 |
| | SA Abdeen et al. 2011 | 12 | -5 | 21 | +12 | 238 | -89 | 329 | +116 |
| | NSGA-II Abdeen et al.2011 | 13 | -4 | 19 | +10 | 244 | -83 | 323 | +111 |
| | Bavota et al. 2013 | 12 | -6 | 21 | +12 | 236 | -91 | 331 | +118 |
| JHotDraw (+-++-) | NSGA-III | 16 | -8 | 37 | +14 | 391 | -83 | 425 | +84 |
| | IBEA | 18 | -6 | 33 | +10 | 404 | -70 | 418 | +77 |
| | MOEA/D | 16 | -8 | 37 | +14 | 391 | -83 | 433 | +92 |
| | SA Abdeen et al. 2011 | 14 | -10 | 42 | +19 | 384 | -90 | 439 | +98 |
| | NSGA-II Abdeen et al.2011 | 15 | -9 | 40 | +17 | 388 | -86 | 435 | +94 |
| | Bavota et al. 2013 | 11 | -13 | 52 | +29 | 378 | -96 | 445 | +104 |
| JDI-Ford (++--+) | NSGA-III | 14 | -4 | 46 | +21 | 301 | -67 | 412 | +76 |
| | IBEA | 14 | -4 | 46 | +21 | 324 | -44 | 422 | +86 |
| | MOEA/D | 16 | -2 | 39 | +16 | 308 | -60 | 391 | +55 |
| | SA Abdeen et al. 2011 | 13 | -5 | 52 | +27 | 297 | -71 | 421 | +85 |
| | NSGA-II Abdeen et al.2011 | 14 | -4 | 48 | +23 | 304 | -64 | 414 | +78 |
| | Bavota et al. 2013 | 13 | -5 | 52 | +27 | 294 | -74 | 424 | +88 |

### 5.5.2 Results for RQ1.2

To answer RQ1.2, we need to assess the correctness/meaningfulness of the suggested remodularization solutions from developers' stand point. To this end we reported the results of our empirical qualitative evaluation in Figure 10 (MP). As reported in Figure 10, the majority of the suggested solutions by NSGA-III improve significantly the structure (RQ1.1) while preserving the semantic coherence much better than all existing approaches. On average, for all of our five studied systems, 88% of proposed operations are considered as semantically feasible and do not generate semantic incoherence by the software engineers. This score is significantly higher than the ones of the NSGA-II and SA approaches having respectively between 51% and 70%, in average of MP scores on the different systems. However, the performance of the IBEA, MOEA/D and Bavota et al. are close to the performance of our NSGA-II approach in terms of semantic preservation with respectively an average of 84%, 83% and 81% of MP. This can be explained by the fact that semantic measures are considered by these approaches.

Thus, our many-objective approach reduces the number of semantic incoherencies when suggesting ROs. To sum up, our approach perform clearly better for semantics preservation with the cost of a slight degradation in structural improvements compared to the other approaches. This slight loss in the structure (RQ1.1) is largely compensated by the significant improvement of the semantic coherence.

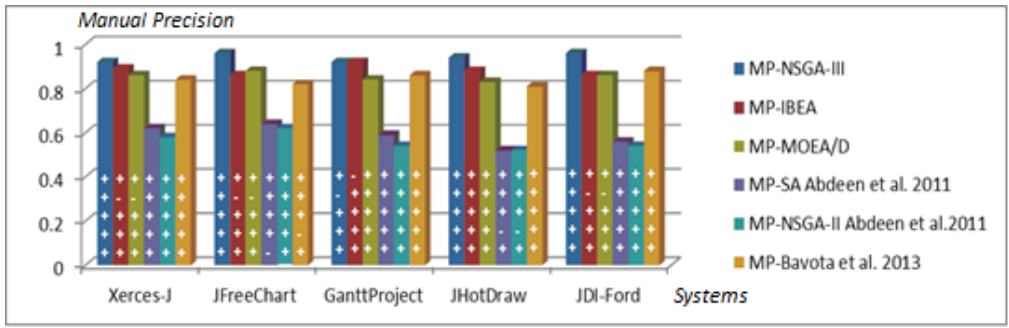

**Fig. 10.** Qualitative evaluation of the remodularization solutions (semantics). A "+" symbol at the $i^{th}$ position in the sequence of signs means that the algorithm metric median value is statistically different from the $i^{th}$ algorithm one. A "-" symbol at the $i^{th}$ position in the sequence of signs means the opposite. sequences of "+" and "-" Signs should be read from top to bottom. The column referring to the system under analysis should be left out of the count in the sequence.

In addition to the empirical evaluation, we automatically evaluate our approach without using the feedback of potential users to give more quantitative evaluation to answer RQ1.2. Thus, we compare the proposed operations with some expected ones. The expected operations are the inverse of those applied manually by the software engineers who participated in our experiments to modify an initial version of JHotDraw. The participants of our experiments did not introduce changes randomly to JHotDraw. We selected the packages that are clearly very well designed based on a manual inspection using also the following quality metrics: number of classes per package, cohesion of the package, number of lines of code per package, average depth of inheritance tree of classes per package, average number of methods per package and coupling of the package. Thus, the expected refactorings are those that can generate the initial version of these modified well-designed packages. The developers considered only move class and split/extract package operations when modifying the original version of JHotDraw. Thus, we are not only considering JHotDraw without a manual inspection of the best well designed packages to modify. The total number of introduced changes is 68. Four developers from the University of Michigan subjects worked together to introduce the changes on the same copy of JHotDraw system. We use Ref-Finder to identify operations that are applied between the program version under analysis and the next version. Figure 11 summarizes our finding. We found that a considerable number of proposed operations by NSGA-III (an average of 85% in terms of precision and recall) are already applied to the next version by software development team comparing to other existing mono-objective and multi-objective remodularization approaches having only less than 65% as precision and recall. However, we found that the remodularization solutions proposed by the other many-objective algorithms IBEA and MOEA/D, and Bavota et al. have close scores to NSGA-III with respectively an average of 80%, 75% and 81% of precision. The same observation is valid for the recall.

In conclusion, our approach produces good remodularization suggestions in terms of structural improvements, semantic coherence, and code changes reduction from the point of view of 1) potential users of our tool and 2) expected operations applied to the next program version.

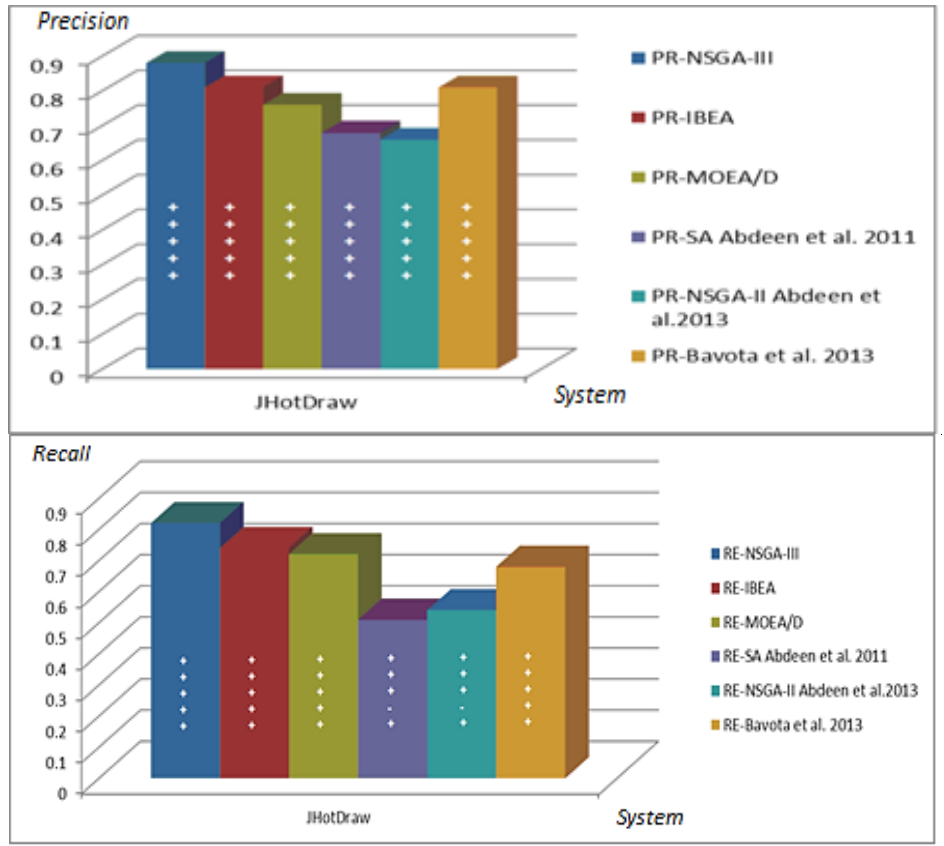

**Fig. 11.** Quantitative evaluation (precision and recall) of the remodularization solutions (semantics). A "+" symbol at the $i^{th}$ position in the sequence of signs means that the algorithm metric median value is statistically different from the $i^{th}$ algorithm one. A "-" symbol at the $i^{th}$ position in the sequence of signs means the opposite. Signs should be read from top to bottom. The column referring to the system under analysis should be left out of the count in the sequence.

*5.5.3 Results for 1.3*

To answer RQ1.3, we evaluate the number of operations (*NO*) suggested by the best remodularization solutions on the different systems. Figure 12 presents the code changes scores (NO) needed to apply the suggested remodularization solutions for each many-objective or multi-objective algorithm. We found that our approach succeeded in suggesting solutions that do not require high code changes (an average of only 64 operations) comparing to other many-objective (IBEA, MOEA/D) and multi-objective (NSGA-II) algorithms having respectively an average of 72, 71 and 79 for all studied systems. We did not compare the number of changes suggested by our proposal with existing work since they are limited to only two type of changes (move class and split packages).

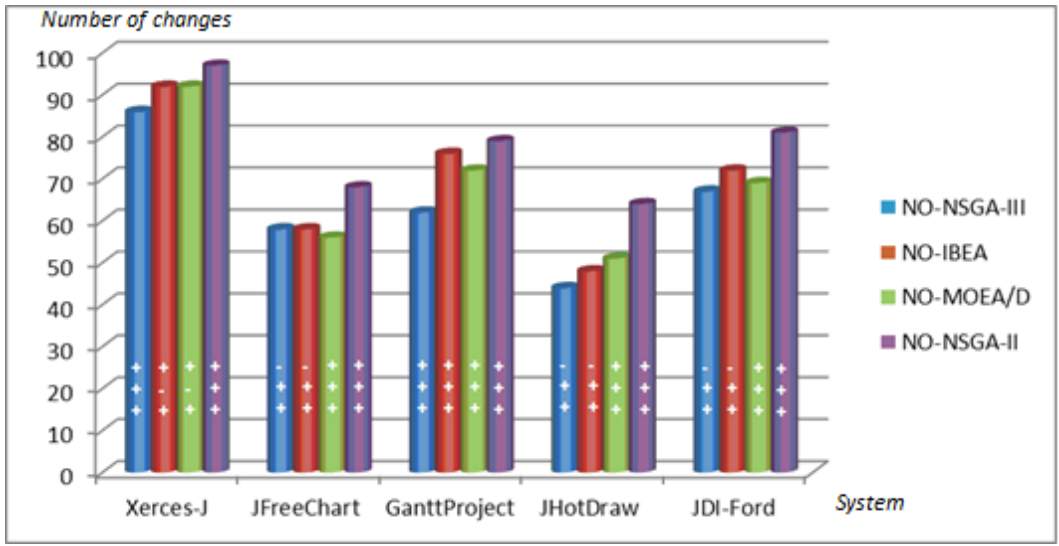

**Fig. 12.** Average number of operations. A "+" symbol at the $i^{th}$ position in the sequence of signs means that the algorithm metric median value is statistically different from the $i^{th}$ algorithm one. A "-" symbol at the $i^{th}$ position in the sequence of signs means the opposite. Signs should be read from top to bottom. The column referring to the system under analysis should be left out of the count in the sequence.

*5.5.4  Results for RQ1.4*

To answer RQ1.4, we evaluated the results of our approach comparing to other approaches that do not use the history of changes. As described in the previous sections, our NSGA-III approach outperforms clearly existing work including Abdeen et al. 2011, Abdeen et al. 2013 and Bavota et al. 2013 that are not based on the use of history of changes. This is a good indication that the recorded operations contribute significantly to provide good solutions. In fact, the use of history of changes is a helper objective to improve the semantic coherence of suggested remodularization solutions. It is also important to note that the SP metrics has the most positive impact on the results of our approach.

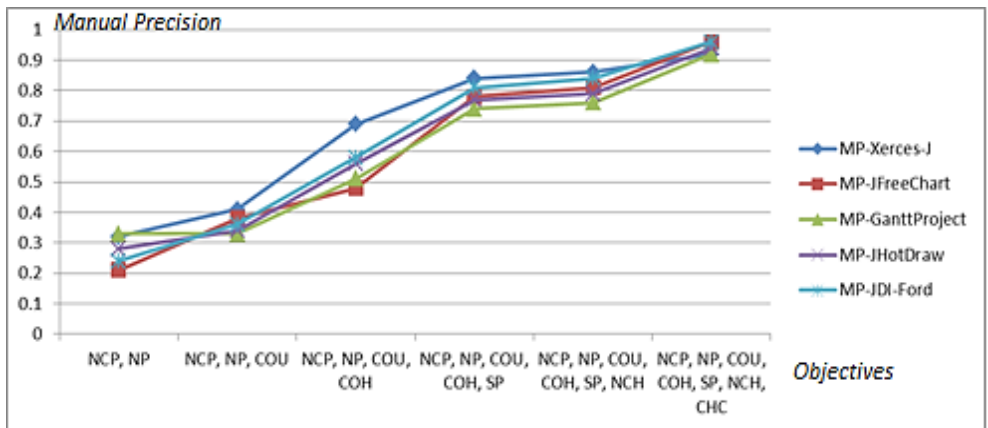

**Fig. 13.** The impact of different combinations of objectives on the remodularization solutions (MP)

We conducted also a more quantitative evaluation to investigate the effects of the use of recorded operations, on the semantic coherence (MP). To this end, we compare the MP score with and without using recorded operations. We present in Figure 13 the results of different combinations of our seven objectives. The order is important only when we considered lower number of objectives (Figure 13) to evaluate the improvement of the solutions quality if a higher number of objectives are used. The first studies on software remodularization used only the structure then the semantics and after that the effort and history of changes. Our work combined them together in one approach using many-objective techniques.

As presented in Figure 13, the best MP scores are obtained when the recorded code changes are considered. Moreover, we found that the optimal remodularization solutions found by our approach are obtained with a considerable percentage of reused operations history (ROP) (more than 75% as shown in Figure 14). Thus, the obtained results support the claim that recorded operations applied in the past are useful to generate coherent and meaningful remodularization solutions.

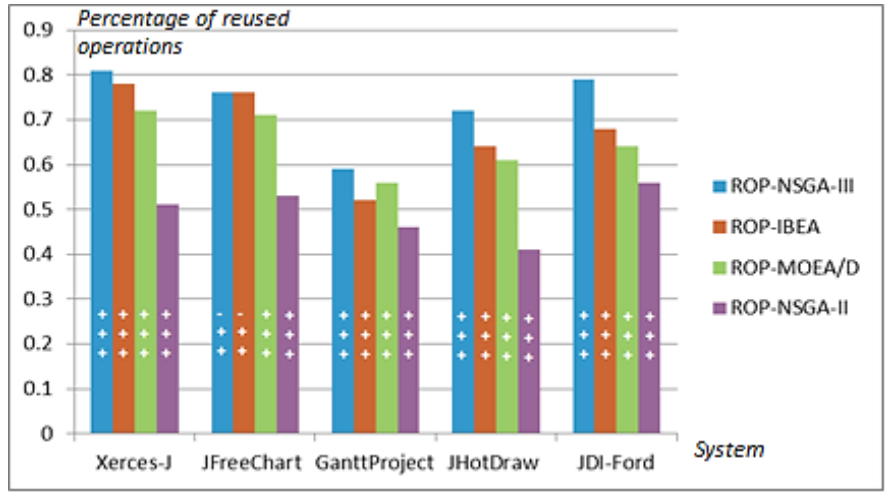

**Fig. 14.** Percentage of recorded operations that are used by the best remodularization. A "+" symbol at the $i^{th}$ position in the sequence of signs means that the algorithm metric median value is statistically different from the $i^{th}$ algorithm one. A "-" symbol at the $i^{th}$ position in the sequence of signs means the opposite solutions. Signs should be read from top to bottom. The column referring to the system under analysis should be left out of the count in the sequence.

*5.5.5   Results for RQ2*

In the previous sections, we compared our NSGA-III proposal with one mono-objective technique and one existing multi-objective technique based on NSGA-II. Thus, we focus on the comparison between our NSGA-III adaption and two other many-objective algorithms IBEA and MOEA/D using the same adaptation. Table 10 shows the median IGD values over 31 independent runs for all algorithms under comparison. All the results were statistically significant on the 31 independent simulations using the Wilcoxon rank sum test with a 99% confidence level (α < 1%). For the 3-objective case, we see that NSGA-III and NSGA-II present similar results, and that NSGA-III provides slightly better results than IBEA and MOEA/D. For the 5-objective case, NSGA-III strictly outperforms NSGA-II and gives similar results to those of the two other multi-objective algorithms. For the 7-objective case, NSGA-III is better than NSGA-II, IBEA and MOEA/D. Additionally, IBEA seems to be slightly better than MOEA/D. It is worth noting that for problems instances with more 3 objectives, NSGA-II performance is dramatically degraded, which is simply denoted by the ~ symbol. The performance of NSGA-III could be explained by the interaction between: (1) Pareto dominance-based selection and (2) reference point-based selection, which is the distinguishing feature of NSGA-III compared to other existing many-objective algorithms.

**Table 10.** Median IGD values on 31 runs (best values are in bold). ~ means a large value that is not interesting to show. The results were statistically significant on 31 independent runs using the Wilcoxon rank sum test with a 99% confidence level (α < 1%).

| System | M | MaxGen | NSGA-III | IBEA | MOEA/D | NSGA-II |
|---|---|---|---|---|---|---|
| Xerces-J | 3 | 250 | **9.861 x 10$^{-4}$** | 9.864 x 10$^{-4}$ | 9.863 x 10$^{-4}$ | 9.862 x 10$^{-4}$ |
|  | 5 | 500 | **7.799 x 10$^{-3}$** | 7.875 x 10$^{-3}$ | 7.878 x 10$^{-3}$ | 8.991 x 10$^{-3}$ |
|  | 7 | 750 | **8.013 x 10$^{-3}$** | 8.372 x 10$^{-3}$ | 8.368 x 10$^{-3}$ | ~ |
| JHotDraw | 3 | 250 | **2.477 x 10$^{-3}$** | 2.478 x 10$^{-3}$ | 2.478 x 10$^{-3}$ | **2.477 x 10$^{-3}$** |
|  | 5 | 500 | **4.193 x 10$^{-3}$** | 4.201 x 10$^{-3}$ | 4.206 x 10$^{-3}$ | 4.533 x 10$^{-3}$ |
|  | 7 | 750 | **5.536 x 10$^{-3}$** | 5.801 x 10$^{-3}$ | 5.796 x 10$^{-3}$ | ~ |
| JFreeChart | 3 | 250 | **3.744 x 10$^{-4}$** | 3.747 x 10$^{-4}$ | 3.746 x 10$^{-4}$ | 3.746 x 10$^{-4}$ |
|  | 5 | 500 | **4.578 x 10$^{-4}$** | 4.602 x 10$^{-4}$ | 4.609 x 10$^{-4}$ | 5.042 x 10$^{-4}$ |

| | 7 | 750 | **6.099 x $10^{-4}$** | 6.208 x $10^{-4}$ | 6.193 x $10^{-4}$ | ~ |
|---|---|---|---|---|---|---|
| GanttProject | 3 | 250 | **5.112 x $10^{-3}$** | 5.115 x $10^{-3}$ | 5.116 x $10^{-3}$ | **5.112 x $10^{-3}$** |
| | 5 | 500 | **6.701 x $10^{-3}$** | 6.802 x $10^{-3}$ | 6.801 x $10^{-3}$ | 6.997 x $10^{-3}$ |
| | 7 | 750 | **7.823 x $10^{-3}$** | 8.068 x $10^{-3}$ | 8.044 x $10^{-3}$ | ~ |
| JDI-Ford | 3 | 250 | **6.229 x $10^{-4}$** | 6.232 x $10^{-4}$ | 6.231 x $10^{-4}$ | 6.231 x $10^{-4}$ |
| | 5 | 500 | **6.608 x $10^{-4}$** | 6.682 x $10^{-4}$ | 6.686 x $10^{-4}$ | 6.887 x $10^{-4}$ |
| | 7 | 750 | **6.984 x $10^{-4}$** | 7.305 x $10^{-4}$ | 7.299 x $10^{-4}$ | ~ |

Figure 15 illustrates the value path plots of all algorithms the 7-objective remodularization problem on JDI-Ford, one of the largest system used in our experiments. Similar observations were made in the remaining systems but are omitted due to space considerations. All quality metrics were normalized between 0 and 1 and all are to be minimized. We observe that NSGA-III and MOEA/D present the best convergence since their non-dominated solution sets are the closest to the ideal point, i.e., the vector composed of 7 zero. However, NSGA-III presents better diversity than all algorithms under comparisons including MOEA/D since its non-dominated solutions have a better spread varying approximately in [0, 0.88] which is not the case for MOEA/D (varying in [0, 0.8]). Besides, MOEA/D seems to have a better convergence than IBEA. However, NSGA-II is unable to progress in terms of convergence as its non-dominated solutions are so far from the ideal vector, and even it diversity is so reduced which may explain the stagnation of its evolutionary process. We conclude that although NSGA-II is the most famous multi-objective algorithm in SBSE, it is not adequate for problems involving over 3 objectives. Based on the results we obtained, it appears that NSGA-III is a very good candidate solution for tackling many-objective SBSE problems.

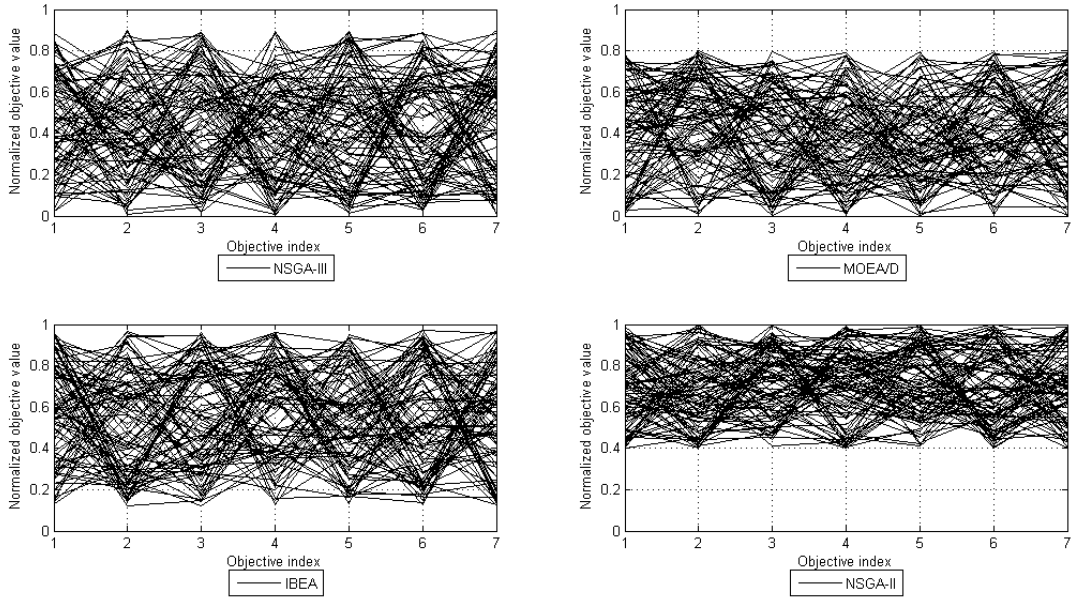

**Fig. 15.** Value path plots of non-dominated solutions obtained by NSGA-III, MOEA/D, IBEA and NSGA-II during the median run of the 7-objective remodularization problem on JDI-Ford.

The results of Figure 15 are not sufficient to show that the 7 objectives are conflicting. Table 11 presents the results of studying the conflict relation between each pair of objectives. In fact, for each experiment (each plot), we execute a mono-objective GA minimizing the objective shown in the line label and we study the behavior of the other objective shown in the column label by recording its values at the beginning (generation 0) and at the end (generation 300) of the evolutionary process. For example, for the first plot (NCP, NP), we observe that the minimization of NCP causes the maximization of NP, thus NCP and NP are conflicting. The opposite phenomenon could be seen for the plot (NP, CHC). Indeed, the minimization of NP makes CHC decreasing too. We conclude that NP and CHC are in support (non-conflicting). To sum up, there are some pairs of objectives that are in conflict and some others that are in support. For this reason, we should verify whether there are some redundant objectives or not [Brockhoff and Zitzler 2006; Saxena et al. 2013], i.e., whether there are some objectives that could be omitted while preserving the Pareto dominance order.

Based on Table 11 results, we draw Table 12 which illustrates for each objective $i$, the objectives that are in conflict with it. We observe from this table that each objective $i$ has a set of conflicting objectives that is different from all other objectives' ones, which means that we cannot omit any objective when comparing between any pair of solutions. Indeed, a redundant objective could appear if two non-conflicting objectives have the same set of conflicting objectives [Brockhoff and Zitzler 2009; Jaimes et al. 2014] which is not the case for our remodularization problem. Thus, we can say that the latter well involves 7 objectives to be optimized simultaneously without omitting anyone of them.

Table 11. Conflict study between objectives.

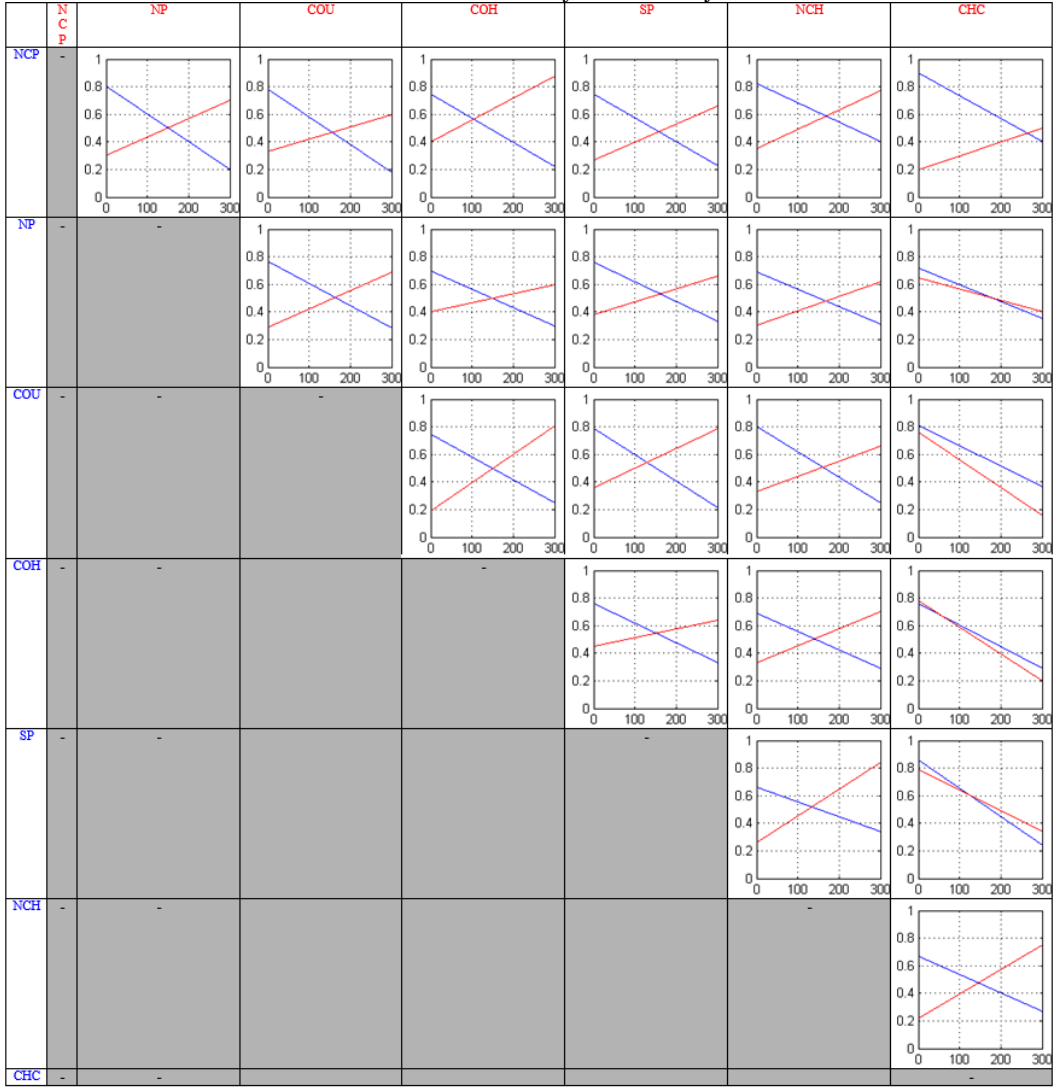

Table 12. Existing conflict between objectives.

| Objective $i$ | Conflicting objectives with $i$ |
|---|---|
| NCP | NP, COU, COH, SP, NCH. |
| NP | COU, NCP, COH, SP, NCH. |
| COU | NCP, NP, COH, SP, NCH. |
| COH | NCP, NP, COU, SP, NCH. |
| SP | NCP, NP, COU, COH, NCH. |
| NCH | NCP, NP, COU, COH, SP, CHC. |
| CHC | NCP, NCH. |

The Wilcoxon rank sum test allows verifying whether the results are statistically different or not. However, it does not give any idea about the difference magnitude. The effect size could be computed by using the Cohen's $d$ statistic [Cohen 1988]. The effect size is considered: (1) small if $0.2 \leq d < 0.5$, (2) medium if $0.5 \leq d < 0.8$, or (3) large if $d \geq 0.8$. For all experiments, we obtained a large difference between NSGA-III/IBEA/MOEA/D results and NSGA-II ones for the cases of 5 and 7 objectives using all the evaluation metrics. The same difference is small for the case of 3 objectives. However, when comparing NSGA-III against MOEA/D and IBEA, we have found the following results: a) On small and medium scale Software systems (JFreeChart and GanttProject) NSGA-III is better that MOEA/D and IBEA on most systems with a medium effect size; b) On large scale Software systems (Xerces-J and JDI-Ford), NSGA-III is better than MOEA/D and IBEA on most systems with a small effect size.

When using optimization techniques, the most time consuming operation is the evaluation step. Thus, we studied the execution time of all many/multi-objective algorithms used in our experiments. Figure 16 shows the evolution of the execution time of the different algorithms on the JDI-Ford system, one of the largest systems in our experiments. The results show that the execution time grows linearly with respect to the number of objectives. It is clear from this figure, that all the algorithms have similar running times for the 3-objective cases. However, for higher number of objectives NSGA-III is faster than IBEA. This observation could be explained by the computational effort required to compute the contribution of each solution in terms of hypervolume. In comparison to MOEA/D, MOEA/D is slightly faster than NSGA-III since it does not make use of non-dominated sorting.

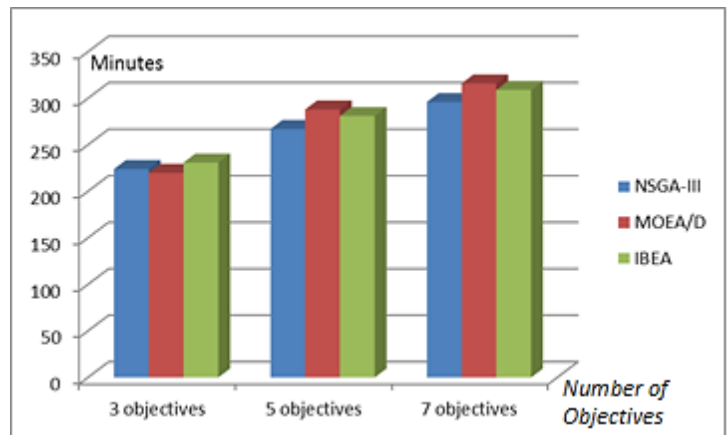

**Fig. 16.** Computational time of the different used many-objective remodularization algorithms.

To further evaluate the scalability of the performance of our many-objective approach on systems of increasing size, we executed our remodularization tool on the Eclipse system without assessing the quality of the results. Eclipse is an open source integrated development environment (IDE) written in Java and widely used to develop applications. We considered four versions of Eclipse that contains more than 3 MLOCs. Figure 17 describes the execution time of our many-objective approach on 4 different versions of Eclipse using the 7 objectives. We believe that an execution time of 8 hours is still acceptable and reasonable even with the considered huge system to remodularize. Developers can execute our tool overnight then evaluate the results and work next day on the new system.

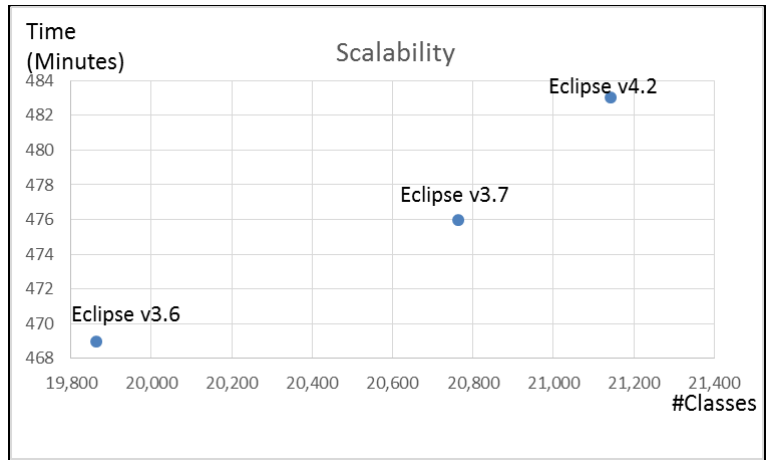
**Fig. 17.** Scalability of our remodularization tool tested on Eclipse.

### 5.5.6 Results for RQ3

We compared the results of our proposal with an existing non-search-based work 0of [Bavota et al. 2013] that eventually relies on the use of coupling and cohesion. In addition, the user needs to give as an input the different packages to restructure. We first note that 0it (like mono-objective approaches also) provides only one remodularization solution, while NSGA-III generate a set of non-dominated solutions. In order to make meaningful comparisons, we select the best solution for NSGA-III using a knee point strategy. For Bavota et al. study, we use the best solution corresponding to the median observation on 31 runs. The results from the 31 runs are depicted in Figures 10 and 11, and Table 9. It can be seen that NSGA-III provides better results than 0 Bavota et al. in all systems as discussed in the previous sections. As described in Figure 18, the NSGA-III outperforms 0it mainly due the use of history of changes as a helper objective for the semantic measures. In addition, Bavota et al. work is limited to few types of operations and do not consider all the operation types supported by our approach.

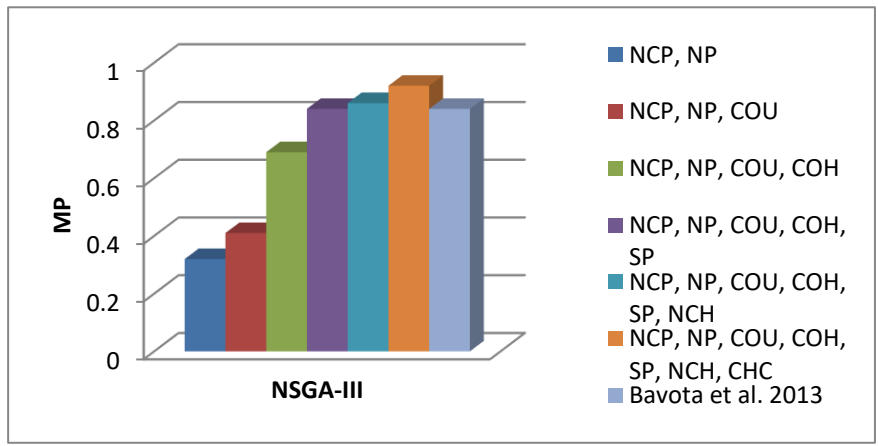
**Fig. 18.** A comparison between Bavota et al. 2013 and NSGA-III based on the qualitative evaluation (MP).

### 5.5.7 Results for RQ4

We asked the software engineers involved in our experiments to evaluate the usefulness of the suggested ROs to apply one by one. In fact, sometimes these operations can improve structure and preserve the semantics but developers will consider them as not useful due to many reasons such as some packages are not used/updated anymore or includes some features that are not important. Figure 19 shows that NSGA-III clearly outperforms existing work by suggesting useful remodularization operations for developers. This is can be explained mainly by the use of the history of recorded changes when suggested remodularization solutions. In fact, the use of the history of changes can help our technique to identify which packages are widely updated. In addition, several recent empirical studies showed that repetitive changes are common during the development of systems [Kim et al. 2013]. We found, in our experiments, that several patterns of changes (applied to different code locations) are repetitive. For example, move methods are in general applied after extract class since several methods are moved to the new created class. Similar observation is valid for extract package and move class. Thus, the use of the history of changes can guide the search for good remodularization solutions based on the reuse of patterns identified from the history of changes.

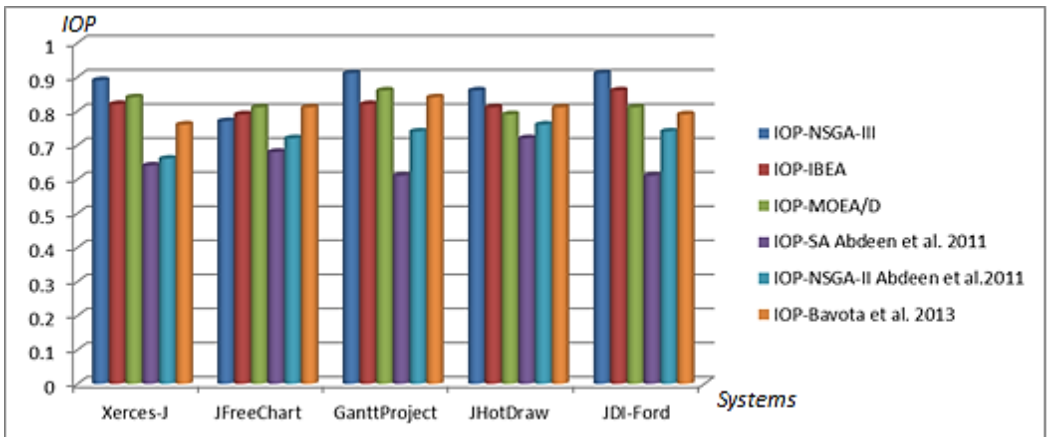

**Fig. 19.** Qualitative evaluation of the suggested remodularization solutions in terms of usefulness.

Another feature that the software engineers, involved in our experiments, found it interesting is the use of several types of ROs. Figure 20 describes the distribution of the operations types used by the best solutions in all the system. It is clear that the three most important ones are move method, move class and extract/split packages. The software engineers found the idea very useful of moving methods between classes located in different packages or extracting a class then moving it to another class instead of moving the whole initial class to a new package. Sometimes, it is enough to move only a method from a class to another class in order to improve the cohesion of a package or decrease coupling between packages. However, existing remodularization work are limited to only two types of operations (move class and split packages).

During the survey, the software engineers confirm that the main limitation related to the use of NSGA-III for software remodularization is the high number of equivalent solutions. However, found the idea of the use of the Knee point as described previously useful to select a good solution. We will investigate in our future work different other techniques to select the region of interest based on the preferences of developers.

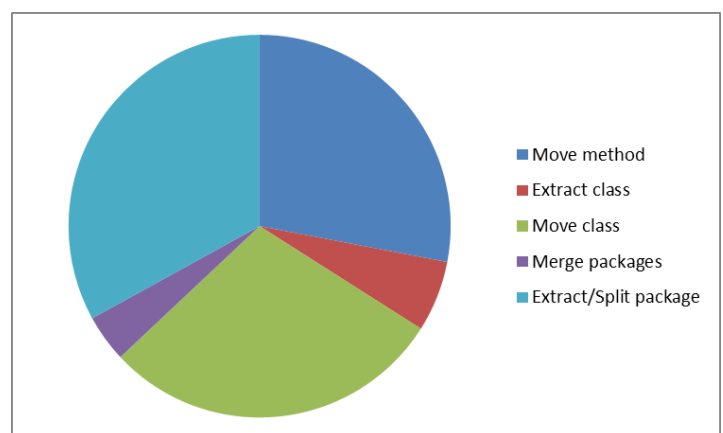

**Fig. 20.** Distribution of the types of suggested remodularization operations

## 6. THREATS TO VALIDITY

We explore in this section the factors that can bias our empirical study. These factors can be classified in three categories: construct, internal and external validity. Construct validity concerns the relation between the theory and the observation. Internal validity concerns possible bias with the results obtained by our proposal. Finally external validity is related to the generalization of observed results outside the sample instances used in the experiment.

In our experiments, construct validity threats are related to the several quantitative measures used in our experiments. To mitigate this threat, we manually inspect and validate the remodularization solutions by a set of experts. Another threat concerns the data about the expected operations of the studied systems. In addition to the documented operations, we are using Ref-Finder which is known to be efficient. Indeed, Ref-Finder was able to detect operations with an average recall of 99% and an average precision of 79%. To ensure the precision, we manually inspect the operations found by Ref-Finder and select only those types considered in our experiments.

We take into consideration the internal threats to validity in the use of stochastic algorithms since our experimental study is performed based on 31 independent simulation runs for each problem instance and the obtained results are statistically analyzed by using the Wilcoxon rank sum test with a 99% confidence level (α = 1%). However, the parameter tuning of the different optimization algorithms used in our experiments creates another internal threat that we need to evaluate in our future work. In fact, parameter tuning of search algorithms is still an open research challenge till today. We have used the trial-and-error method which one of the most used ones [Eiben and Smit 2011]. However, the use of ANOVA-based technique [Miller and Rupert 1986] could be another interesting direction from the viewpoint of the sensitivity to the parameter values. Another threat is related to the order of applying the objectives that my influences the outcome of the search. We are planning to investigate the impact of the order of objectives on the results by evaluating several combinations. In addition, the weights used for the semantic functions are selected manually and further experiments are required to study the impact of the variation of these weights on the quality of the results. Another threat is related to our experiments to show how the execution time is related to the size of the system in terms of number of classes. The number of classes of a system can be not enough to have a strong conclusion especially when other factors such as the density of the interdependency graph probably will influence execution time more than the number of classes in the system. In addition, it is may not be enough to show with 3 version of Eclipse if the execution time is a linear or non-linear.

We identify other three threats to internal validity: selection, learning and fatigue, and diffusion. Another internal threat is related to the problem of isomorphic solutions since inferring goodness based on an objective function can sometimes be misleading. To this end, we manually validated the solution as described in the experiments and we found correlation between the fitness function values and the success metrics (different from the fitness functions and correspond to the developers opinion after manually validating some solutions) used in the experiments. However, we cannot generalize our correlation results since it was limited to few solutions and the problem of isomorphic solutions is out of the scope of this paper. In addition, it is challenging to address this problem especially for the case of many-objective optimization where a high number of solutions are generated. Thus, we plan in our future work to study the correlation through extensive empirical studies between the improvements of the fitness function values and the quality of solutions validated manually by experts on several software engineering problems. We have also addressed in our very recent work [Ben Said et al. 2010] the problem of subjective fitness function or when the fitness function is difficult to define. To this end, we proposed an approach based on the use of Machine Learning (ANNs) to evaluate solutions based on a training set of solutions evaluated manually by experts. This can lead to more realistic fitness functions that can describe how solutions are evaluated by software engineers for specific problems such as software refactoring. We will extend that work for the remodularization problem. An additional internal threat is that our approach is limited to the use of static metrics analysis. However, our approach is generic thus additional objectives/metrics and inputs can be easily added to extend our algorithm. To this end, we are planning to use dynamic analysis techniques to evaluate the system after remodularization and evaluate the impact of suggested operations on the dynamic/runtime relations.

For the selection threat, the subject diversity in terms of profile and experience could affect our study. First, all subjects were volunteers. We also mitigated the selection threat by giving written guidelines and examples of operations already evaluated with arguments and justification. Additionally, each group of subjects evaluated different operations from different systems using different techniques/algorithms.

Randomization also helps to prevent the learning and fatigue threats. For the fatigue threat, specifically, we did not limit the time to fill the questionnaire. Consequently, we sent the questionnaires to the subjects by email and gave them enough time to complete the tasks. Finally, only ten operations per system were randomly picked for the evaluation.

Diffusion threat is limited in our study because most of the subjects are geographically located in a university and a company, and the majority does not know each other. For the ones who are in the same location, they were instructed not to share information about the experience before a certain date.

To ensure the heterogeneity of subjects and their differences, we took special care to diversify them in terms of professional status, university/company affiliations, gender, and years of experience. In addition, we organized subjects into balanced groups. This has been said, we plan to test our tool with Java development companies, to draw better conclusions. Moreover, the automatic evaluation is also a way to limit the threats related to subjects as it helps to ensure that our approach is efficient and useful in practice. Indeed, we compare our suggested operations with the expected ones that are already applied to the next releases and detected using Ref-Finder.

External validity refers to the generalizability of our findings. In this study, we performed our experiments on five different systems belonging to different domains and with different sizes. However, we cannot assert that our results can be generalized to other applications, other programming languages, and to other practitioners. Future replications of this study are necessary to confirm the generalizability of our findings. The participants considered in our experiments are not the original developers of the open source systems thus some of their evaluations of the remodularization solutions could be not very accurate since there are, sometimes, good reasons for the design and implementation choices made and this can be mainly determined by the original developers. However, this not the case for the Ford project since some of the original developers of the system participated in our experiments. In addition, the number of participants in our experiments is limited. We are planning to integrate additional original developers from these open source projects to evaluate the detected code smells as part of our future work.

## 7. RELATED WORK

In the last two decades, a large number of research works have been proposed in the literature to support (semi-)automatic approaches to help software engineers maintain and extend existing systems. In fact, several studies addressed the problem of clustering to find the best decomposition of a system in terms of modules and not improving existing modularizations. In this section, we focus on existing remodularization studies based on refactoring.

[Wiggerts 1997] provides the theoretical background for the application of cluster analysis in systems remodularization. He discusses on how to establish similarity criteria between the entities to cluster and provide the summary of possible clustering algorithms to use in system remodularization. Later, [Anquetil and Lethbridge 1999] use cohesion and coupling of modules within a decomposition to evaluate its quality. They tested some of the algorithms proposed by Wiggerts and compared their strengths and weaknesses when applied to system remodularization. [Maqbool et al. 2007] focus on the application of hierarchical clustering in the context of software architecture recovery and modularization. They investigate the measures to use in this domain, categorizing various similarity and distance measures into families according to their characteristics. A more recent work by [Shtern and Azerpos 2009] introduced a formal description template for software clustering algorithms. Based on this template, they proposed a novel method for the selection of a software clustering algorithm for specific needs, as well as a method for software clustering algorithm improvement. The underlying idea of these approaches is to 1) group in a module highly cohesive source code components, where the cohesiveness is measured in terms of intra-module links; and 2) reduce the coupling between modules, where the coupling is measured in terms of inter-module dependencies. The vast majority of clustering-based approaches aim at modularizing/decomposing software systems from scratch using only structural measures. However, the goal of our approach aims at assisting software maintainers in the task of improving the quality of existing packages structure while maintaining the semantic coherence of the original design structure.

There have been several developments in search-based approaches to support the automation of software modularization. [Mancoridis et al. 1998] was the first search-based approach to address the problem of software modularization using a single-objective approach. Their idea to identify the modularization of a software system is based on the use of the hill-climbing search heuristic to maximize cohesion and minimize coupling. The same technique has been also used in [Mitchell and Mancoridis 2006] and [Mitchell and Mancoridis 2008] where the authors present Bunch, a tool supporting automatic system decomposition. Subsystem decomposition is performed by Bunch by partitioning a graph of entities and relations in a given source code. To evaluate the quality of the graph partition, a fitness function is used to find the trade-off between interconnectivity (i.e., dependencies between the modules of two distinct subsystems) and intra-connectivity (i.e., dependencies between the modules of the same subsystem), to found out a satisfactory solution. [Harman et al. 2002] use a genetic algorithm to improve the subsystem decomposition of a software system. The fitness function to maximize is defined using a combination of quality metrics, e.g., coupling, cohesion, and complexity. Similarly, [Seng et al. 2005] treated the remodularization task as a single-objective optimization problem using genetic algorithm. The goal is to develop a methodology for object-oriented systems that, starting from an existing subsystem decomposition, determines a decomposition with better metric values and fewer violations of design principles. [Abdeen et al. 2009] proposed a heuristic search-based approach for automatically optimizing (i.e., reducing) the dependencies between packages of a software system using simulated annealing. Their optimization technique, is based on moving classes between the original packages. Taking inspiration from our previous work [Ouni et al. 2012a], [Abdeen et al. 2013] have recently extended their initial work to consider the remodularization task as a multi-objective optimization problem to improve existing packages structure while minimizing the modification amount on the original design. Using NSGA-II, this optimization approach aims at increasing the cohesion and reducing the coupling and cyclic connectivity of packages, by modifying as less as possible the existing packages organization. [Praditwong et al. 2011] have recently formulated the software clustering problem as a multi-objective optimization problem. Their work aim at maximizing the modularization quality measurement, minimizing the inter-package dependencies, increasing intra-package dependencies, maximizing the number of clusters having similar sizes and minimizing the number of isolated clusters.

Most of the remodularization approaches in the literature are based on information derived only from structural metrics to modularize/restructure the original package organization. However, this is not enough to produce a semantically coherent design. The first attempt that addresses this problem was by [Bavota et al. 2010] who proposed an automated, single-objective, approach to split an existing package into smaller but more cohesive ones. The proposed approach analyzes the structural and semantic relationships between classes in a package identifying chains of strongly related classes. The identified chains are used to define new packages with higher cohesion than the original package. This work has been extended in [Bavota et al. 2012], to propose an interactive multi-objective remodularization approach. The proposed Interactive Genetic Algorithms (IGAs) aims at integrating developer's knowledge in a remodularization task. Specifically, the proposed algorithm uses a fitness composed of automatically-evaluated factors (accounting for the modularization quality achieved by the solution) and a human-evaluated factor, penalizing cases where the way remodularization places components into modules is considered meaningless by the developer. One of the limitations of this approach is that, in each generation of the remodularization process, end users should analyze the suggested solution, class-by-class and package-by-package, and provide their feedback. User feedback can be either about classes which should stay together, or not, and/or about small/isolated clusters. This is not always profitable when we deal with industrial size software projects, and it need expert users to suitably drive the optimization process.

The semantic meaningfulness of the recommended restructuration is a fundamental issue when automatically modifying a software design. The first attempt to integrate the semantic coherence of when automatically modifying the software design was in [Ouni et al. 2012b]. Similarly, our remodularization approach use the combination of semantic and structural information captured in the package and class levels to suggest more meaningful remodularization and better decide how to group together, split, or move, (or not) certain code elements. Furthermore, while automatic remodularization approaches proved to be very effective to increase cohesiveness and reduce coupling of software modules, they do not take into account the history of changes that provide a lot of information that is very useful in automating many software maintenance tasks. One of the characteristics of our approach is that it exploits the change history that represents an effective way to produce more meaningful remodularization. Another issue is that the majority of existing remodularization approaches considers only moving classes or grouping/splitting packages; however, none considered move methods/fields among classes in different packages. Hence, sometimes, it is enough to move only a method or a field between two classes in two different packages to reduce the dependency between them.

8. **CONCLUSIONS AND FUTURE WORK**

In this paper we introduced a new scalable search-based software engineering approach for software remodularization based on NSGA-III. This paper represents the first real-world application of NSGA-III and the first scalable work that supports the use of 7 objectives to address a software engineering problem. We address several challenges of existing software remodularization techniques that are limited to mainly the use of coupling and cohesion, and few types of operations (move class and split package). Our proposal aims at finding the remodularization solution that improve the structure of packages by optimizing some metrics such as number of classes per package, number of packages, coupling and cohesion; improve the semantic coherence of the restructured program; minimize code changes; and maximize the consistency with development change history.

We evaluated our approach on four open source systems and one industrial system provided by our industrial partner Ford Motor Company. We report the results on the efficiency and effectiveness of our approach, compared to the state of the art remodularization approaches. Our results indicate that our approach significantly outperforms, in average, existing approaches based on a quantitative and qualitative evaluation. All the results were statistically significant on the 31 independent simulations using the Wilcoxon rank sum test with a 99% confidence level ($α < 1\%$) where more than 92% of code-smells were fixed on the different open source systems.

As part of the future work, we plan to work on adapting NSGA-III to additional software engineering problems and we will perform more comparative studies on larger open source systems. Furthermore, we will investigate the impact of different parameter settings on the quality of our results. Nevertheless, this extensive study has shown a direction using NSGA-III to handle as many as 7 objectives in the context of solving software engineering problems and would remain as one of the first studies in which such a large number of objectives have been considered. Weighting certain objectives can be an interesting future work direction to integrate the preferences of developers during the search process. We are also planning to consider only the remodularization of the modified packages after the last release based on analysis of the history of the code changes. Finally, we plan to extend the use of our modularization approach by additional experts to generalize the obtained results.